\documentclass[aps,prd,floatfix,amsfonts,amssymb,amsmath,twocolumn,notitlepage,nofootinbib,groupedaddress,superscriptaddress]{revtex4-1}
\usepackage{graphicx}
\usepackage{slashed}
\usepackage[T1]{fontenc}
\usepackage{color}
\usepackage{hyperref}
\usepackage{times}

\DeclareMathOperator{\e}{\operatorname{e}}

\begin{document}
\title{Thermodynamic phases and mesonic fluctuations in a chiral nucleon-meson model}
\author{Matthias Drews}
	\affiliation{ECT*, Villa Tambosi, I-38123 Villazzano (Trento), Italy}
	\affiliation{Physik Department, Technische Universit\"{a}t M\"{u}nchen, D-85747 Garching, Germany}
\author{Thomas Hell}
	\affiliation{ECT*, Villa Tambosi, I-38123 Villazzano (Trento), Italy}
	\affiliation{Physik Department, Technische Universit\"{a}t M\"{u}nchen, D-85747 Garching, Germany}
\author{Bertram Klein}
	\affiliation{Physik Department, Technische Universit\"{a}t M\"{u}nchen, D-85747 Garching, Germany}
\author{Wolfram Weise}
	\affiliation{ECT*, Villa Tambosi, I-38123 Villazzano (Trento), Italy}
	\affiliation{Physik Department, Technische Universit\"{a}t M\"{u}nchen, D-85747 Garching, Germany}
\date{\today}

\begin{abstract}
Studies of the QCD phase diagram must properly include nucleonic degrees of freedom and their thermodynamics in the range of baryon chemical potentials characteristic of nuclear matter. A useful framework for incorporating relevant nuclear physics constraints in this context is a chiral nucleon-meson
effective Lagrangian. In the present paper, such a chiral nucleon-meson model is extended with systematic inclusion of mesonic fluctuations using the functional renormalization group approach. The resulting description of the nuclear liquid-gas phase transition shows a remarkable agreement with three-loop calculations based on in-medium chiral effective field theory. No signs of a chiral first-order phase transition and its critical endpoint are found in the region of applicability of the model, i.e. up to twice the density of normal nuclear matter and at temperatures $T \lesssim 100$~MeV. Fluctuations close to the critical point of the first-order liquid-gas transition are also examined with a detailed study of chiral and baryon number susceptibilities.
\end{abstract}

\maketitle

\begin{section}{Introduction}
A persistently interesting problem in quantum chromodynamics (QCD) at finite temperature and baryon chemical potential is the exploration of the phase diagram and the quest for the existence of a critical endpoint associated with a first-order chiral phase transition. For large quark chemical potentials, $\mu_q$, \textit{ab initio} lattice QCD calculations are restricted by the sign problem to \mbox{$\mu_q/T < 1$}~\cite{Philipsen2005qcd,Schmidt2006lattice,Philipsen2008status,de_Forcrand2009simulating}. A lattice simulation at imaginary chemical potential for relatively large quark masses indicates a shrinking critical surface of the first-order critical region~\cite{forcrand2008chiral} which would exclude a critical endpoint for physical quark masses. 

Given that large baryon chemical potentials at low temperatures are not directly accessible to lattice QCD computations, this region has therefore been subject to a variety of model investigations (see, e.g., \cite{fukushima2013phase} for a recent survey). In the Nambu--Jona-Lasinio (NJL) model and its Polyakov-loop-extended (PNJL) version~\cite{fukushima2004chiral,roener2007polyakov,ratti2006thermodynamics} the existence of the critical endpoint of a possible chiral first-order phase transition depends sensitively on the 't~Hooft coupling representing the axial $U(1)$ anomaly, and on the strength of vector current interactions. This was demonstrated in the NJL model \cite{klimt1990chiral,sasaki2007quark,kitazawa2002chiral} and for the local PNJL model~\cite{fukushima2008phase,bratovic2013role} as well as its nonlocal versions~\cite{hell2013impact}. The vector interaction, in particular, is shown to contribute crucially to the curvature of the critical surface~\cite{fukushima2008critical}. 

When fluctuations beyond the mean-field approximation are included, the critical endpoint appears generally at smaller temperatures and larger chemical potentials. This effect was observed in the Polyakov-loop-extended quark-meson (PQM) model~\cite{schaefer2012qcd} with matter backreactions taken into account~\cite{schaefer2007phase}, and within a functional renormalization group approach starting from such a model~\cite{Skokov2010Meson,Skokov2011Quark,Herbst2011Phase,Herbst2013Phase}. In the latter case, no critical endpoint is found. Recent studies using Dyson-Schwinger equations find, in contrast, a critical endpoint of the chiral phase transition at a large critical temperature \mbox{$T=100\operatorname{ MeV}$}, and critical quark chemical potential in the range \mbox{$\mu_q=170\text{--}190\operatorname{ MeV}$} \cite{fischer2012propagators,fischer2013polyakov}. Alternative scenarios, such as the possibility of spatially inhomogeneous quark matter, are also discussed in the recent literature \cite{nickel_how_2009,ooguri_spatially_2011,fukushima_spatial_2013}.

The PNJL and PQM models work with quarks as quasiparticles. While color-nonsinglet degrees of freedom are properly suppressed in the hadronic sector of the phase diagram, color-singlet three-quark configurations are not confined in localized clusters to form baryons. Hence, important physics is missing at low temperatures $T$ and at baryon chemical potentials around and below the nucleon mass, \mbox{$\mu~\equiv~\mu_B~\lesssim~1~\text{GeV}$}. In this regime the phase diagram of strongly interacting matter is governed by correlated nucleons (rather than quarks) and by the coexistence of nuclear Fermi liquid and gas phases. A model dealing with this sector of the phase diagram should therefore minimally fulfill the following conditions: it should incorporate the spontaneous breaking of chiral symmetry characteristic of low-energy QCD; it should be based on nucleons and pions as relevant degrees of freedom (plus ingredients such as heavier effective boson fields to account for short-distance dynamics of the nucleon-nucleon interaction); and it should be consistent with well-known nuclear physics facts and constraints.

The thermodynamics of such a chiral nucleon-meson model~\cite{berges2003quark} has recently been studied in mean-field approximation~\cite{floerchinger2012chemical}. No first-order chiral phase transition was found within the region of applicability of this model.
In particular, it turns out~\cite{floerchinger2012chemical} that the chemical freeze-out trajectory in the $T$-$\mu$ plane, deduced from a resonance-gas analysis of heavy-ion collision data~\cite{Braun-Munzinger2004Chemical} is not related to a chiral transition at chemical potentials \mbox{$\mu\sim 700\text{--}900\text{ MeV}$}. This differs from the situation at vanishing $\mu$ where such a connection between freeze-out and chiral crossover around \mbox{$T\simeq 170\text{ MeV}$} is suggestive. Lattice QCD computations also indicate that the chiral transition and chemical freeze-out curves tend to separate as the chemical potential increases~\cite{Kaczmarek2011Phase,Karsch2012Determination}.

The present work focuses on the chiral nucleon-meson model~\cite{floerchinger2012chemical, berges2003quark} with inclusion of mesonic fluctuations beyond the mean-field approximation using the functional renormalization group (FRG) approach~\cite{wetterich1993exact}. The paper is organized as follows: first  we present the nucleon-meson model and show how fluctuations can be included. The nuclear liquid-gas transition is studied and compared to results from chiral effective field  theory ($\chi$EFT) applied to the nuclear many-body problem
~\cite{fiorilla2012chiral,fiorilla2012nuclear}. Then the thermodynamics of the chiral condensate and the question of a critical endpoint for a chiral first-order transition are studied. To anticipate one of the results: we find no such critical endpoint for temperatures \mbox{$T\lesssim 100\operatorname{ MeV}$} and baryon chemical potentials \mbox{$\mu\lesssim 1\operatorname{ GeV}$}.
Finally, we examine chiral and baryon number susceptibilities in the context of the present model.
\end{section}

\begin{section}{Chiral nucleon-meson model}
We begin with a brief description of the chiral nucleon-meson model used in Ref.~\cite{floerchinger2012chemical}. The degrees of freedom at work in baryonic matter at densities around the nuclear liquid-gas phase transition are nucleons and pions, with their dynamics governed by the spontaneously broken chiral symmetry of low-energy QCD.  Chiral symmetry is realized at the level of the effective Lagrangian in the form of a generalized linear sigma model. The nucleon mass is generated by the expectation value of a scalar field $\sigma$. The $\sigma$ and pion fields are combined in a four-component field, \mbox{$\phi=(\sigma,\boldsymbol\pi)$}, that transforms under the chiral group \mbox{$\operatorname{SO}(4)\cong\operatorname{SU}(2)_L\times\operatorname{SU}(2)_R$}. Its invariant square is defined as:
\begin{align}
	\rho=\frac 12\phi^\dagger\phi=\frac 12(\sigma^2+\boldsymbol\pi^2). 
\end{align}
Protons and neutrons are combined in the isospin doublet Dirac field \mbox{$\psi=(\psi_p,\psi_n)^T$}.
As an additional important ingredient, the repulsive short-range nucleon-nucleon force is conveniently modeled in terms of a four-fermion vector interaction proportional to \mbox{$(\bar\psi\gamma_\mu\psi)\,(\bar\psi\gamma^\mu\psi)$}. When bosonized by means of a Hubbard-Stratonovich transformation, this short-distance repulsion can be thought of as mediated by a vector field, $\omega_\mu$, as in the time-honored Walecka model \cite{walecka1974theory}, but now in a framework that explicitly incorporates chiral symmetry. 

In summary, the Lagrangian (written in Minkowski space-time) of the chiral nucleon-meson model reads
\begin{align}\label{eq:Lagrangian}
	\begin{aligned}
		\mathcal L&=\bar\psi\Big[i\gamma_\mu\partial^\mu-g_s(\sigma+i\gamma_5\,\boldsymbol\tau\cdot\boldsymbol\pi)-g_v \gamma_\mu\omega^\mu\Big]\psi \\
		&\quad+\tfrac 12\partial_\mu\sigma\,\partial^\mu\sigma+\tfrac 12\partial_\mu\boldsymbol\pi\cdot\partial^\mu\boldsymbol\pi - {\cal U}(\boldsymbol\pi,\sigma) \\
		&\quad-\tfrac 14 F_{\mu\nu}F^{\mu\nu} + \tfrac 12m_v^2\,\omega_\mu\omega^\mu\,,
	\end{aligned}
\end{align}
with the field tensor $F_{\mu\nu} = \partial_\mu\omega_\nu - \partial_\nu\omega_\mu$. \mbox{$\boldsymbol\tau=(\tau_1,\,\tau_2,\,\tau_3)$} are Pauli matrices in isospin space. The parameters of the model are the (pseudo)scalar and vector couplings $g_s$ and $g_v$, respectively, and the mass $m_v$ of the vector boson. The potential ${\cal U}(\boldsymbol\pi,\sigma)$ has a chirally invariant piece, ${\cal U}_0(\rho)$, and a term linear in $\sigma$ that breaks chiral symmetry explicitly, 
\begin{align}
	\mathcal U(\boldsymbol\pi,\sigma)  = \mathcal U_0(\rho) - m_\pi^2\,f_\pi\,\sigma\,,
\end{align}
involving  the pion mass, \mbox{$m_\pi=139\operatorname{ MeV}$}, together with the pion decay constant, \mbox{$f_\pi = 93\operatorname{ MeV}$}.

The treatment of the equilibrium thermodynamics based on this model Lagrangian involves the following standard steps. First, the action in Minkowski space, $S_{\text M} = \int d^4x \,{\cal L}$, is rewritten in Euclidean space-time with \mbox{$t\equiv x^0 \rightarrow -i\tau$}. The time integral $\int dt$ is replaced by $-i\int_0^\beta d\tau$ with the inverse temperature \mbox{$\beta = 1/T$}. Periodic (antiperiodic) boundary conditions apply to bosonic (fermionic) fields. A nonzero baryon chemical potential, $\mu$, is introduced by adding the term \mbox{$\Delta S_{\text E}(\mu) = -\beta\mu B$} to the Euclidean action $S_{\text E}$, with the baryon number \mbox{$B = \int d^3x \,\psi^\dagger \psi$}.

The aim is now, as in Ref.~\cite{floerchinger2012chemical}, to construct an effective potential $U$ that incorporates quantum and thermal fluctuations. So far, the fields are space-time dependent. In the mean-field approximation, they are replaced by their space-time-independent averaged values. The only fields that can acquire nonzero expectation values are the scalar field $\sigma(x)$, representing the chiral (quark) condensate in the hadronic phase of QCD with spontaneously broken chiral symmetry, and the time component $\omega_0(x)$ of the vector field $\omega_\mu(x)$ linked to the baryon density as its source. For convenience, the mean-field values of $\sigma(x)$ and $\omega_0(x)$ are again denoted by $\sigma$ and $\omega_0$, respectively. The spatial components of the $\omega$ field vanish in the mean-field approximation in order to preserve the rotational symmetry of the vacuum. The expectation value of the pion field vanishes assuming that there is no pion condensate. For constant fields in space-time the minimum of the effective action $\Gamma$ corresponds to the minimum of the effective potential, \mbox{$U = (T/V)\Gamma$}, expressed as a function of the expectation values of the boson fields.

The grand-canonical potential is obtained by minimizing the effective potential as a function of $\sigma$ and $\omega_0$ for a given temperature $T$ and baryon chemical potential $\mu$. The fields at the minimum of the effective potential are denoted as $\bar{\sigma}  (\mu,T)$ and $\bar{\omega}_0(\mu,T)$. The pressure $P$, the baryon density $n$  and the energy density $\cal{E}$ can then be calculated from \mbox{$U(\mu,T) \equiv U\Big(\bar{\sigma}(\mu,T),\bar{\omega}_0(\mu,T);\mu,T\Big) = {\cal E} - Ts - \mu n$} as follows:
\begin{align}
	\begin{gathered}
		P = - U(\mu,T)\,, \qquad n = - {\partial U(\mu,T)\over \partial \mu}\,, \\
	       \mathcal E  = U(\mu,T) + \mu n - T {\partial U(\mu,T)\over \partial T}\,.
       \end{gathered}
\end{align} 
A detailed construction of the effective potential $U(\sigma, \omega_0; T, \mu)$ (starting from the microscopic action and potential ${\cal U}$), is practically not feasible. However, for the thermodynamics derived from the model Lagrangian \eqref{eq:Lagrangian}, only the difference between effective potentials,
\begin{align}\label{eq:action_diff}
	U(\sigma,\omega_0; T,\mu)-U(\sigma,\omega_0; T=0,\mu=\mu_c)\,,
\end{align}
is of interest~\cite{floerchinger2012chemical}. Here $\mu_c$ is the baryon chemical potential at the point where the nuclear liquid-gas phase transition occurs at vanishing temperature. This chemical potential is the difference of the nucleon mass, $m_N$, and binding energy, $B$, in nuclear matter at equilibrium: \mbox{$\mu_c= m_N  - B = (939-16)\text{ MeV}=923\text{ MeV}$}. 

In practice, the effective potential $U$ is parametrized such as to reproduce empirical nuclear physics data at $T=0$ and \mbox{$\mu=\mu_c$}. The chiral symmetric part of the potential is expanded up to a sufficiently high order $N_{\text{max}}$ in the chiral field $\rho = \frac 12(\sigma^2 + \boldsymbol\pi^2)$ (reduced by setting $\boldsymbol\pi = 0$) around its vacuum value, $\rho_0=\frac 12f_\pi^2$: 
\begin{align}
	\begin{aligned}
		U(\sigma,\omega_0)&=-m_\pi^2\, f_\pi(\sigma-f_\pi)+m_\pi^2(\rho-\rho_0) \\
		&\quad+\sum_{n=2}^{N_{\text{max}}}\frac{a_n}{n!}(\rho-\rho_0)^n - \frac 12m_v^2\,\omega_0^2\,.
	\end{aligned}
\end{align}
The coefficient of the term linear in $\rho - \rho_0$ is fixed by the physical pion mass. The explicit symmetry breaking term linear in $\sigma$ fixes the vacuum expectation value of $\sigma$ to $f_\pi$. A constant has been subtracted to achieve a vanishing vacuum pressure, 
\begin{align}
	P_{\text{vac}}=-U(\sigma = f_\pi,\omega_0 = 0)=0\,.
\end{align}
When the nucleons are integrated out the mean-field effective potential takes the form
\begin{align}\label{eq:UMF}
	U_{\text{MF}}&=U(\sigma,\omega_0)+4U_{\text{N}}\,,
\end{align}
where 
\begin{align}
	\begin{aligned}
		U_{\text{N}} &= -\int\frac{d^3p}{(2\pi)^3}\,T\ln\big[1+\e^{-\beta(E_N(p)-\mu_{\text{eff}})}\big] \\
		\quad&-\int\frac{d^3p}{(2\pi)^3}\,T\ln\big[1+\e^{-\beta(E_N(p)+\mu_{\text{eff}})}\big]\,
	\end{aligned}
\end{align}
is the effective potential of (relativistic) nucleon quasiparticles with \mbox{$E_N(p)=\sqrt{p^2+m_{\text{eff}}^2}$}. The prefactor of four in Eq.~\eqref{eq:UMF} accounts for spin and isospin degeneracies. The effective nucleon quasiparticle mass and chemical potential are given as
\begin{align}
	m_{\text{eff}}= g_s\,\sigma\,, \qquad \mu_{\text{eff}}=\mu-g_v\,\omega_0\,.
\end{align}
The presence of the  background vector field shifts (reduces) the baryon chemical potential. 

For a given temperature $T$ and chemical potential $\mu$, the mean-field effective potential (\ref{eq:UMF})
is minimized with respect to $\sigma$ and $\omega_0$,
\begin{align}
	\left.{\partial U_{\text{MF}}\over\partial\sigma}\right|_{\sigma=\bar\sigma,\,\omega_0=\bar\omega_0}=0\,,
	\qquad \left.{\partial U_{\text{MF}}\over\partial\omega_0}\right|_{\sigma=\bar\sigma,\,\omega_0=\bar\omega_0}=0\,.
\end{align}
Minimization with respect to $\omega_0$ gives the self-consistent equation
\begin{align}
	\bar\omega_0 = {g_v\over m_v^2} n(T, \mu-g_v\bar\omega_0)\,,
\end{align}
where the baryon density $n$ is determined by
\begin{align}
	n(T,\mu-g_v\omega_0) = -4\, {\partial \over \partial\mu} U_{\text{N}}(T,\mu-g_v\omega_0)\,.
\end{align}
At this level, the vector coupling $g_v$ and the mass of the $\omega$ field are not independent. 
%After a field redefinition $\omega_0\rightarrow \omega_0/m_v$, 
Only their ratio $g_v/m_v$ appears in the shifted chemical potential that enters the mean-field equations. 
A parametrization of the effective potential with $N_{\text{max}}=4$ that is consistent with nuclear physics constraints is the one chosen in~\cite{floerchinger2012chemical}
\begin{gather}\label{eq:parameterizationMF}
	\begin{gathered}
		g_s=\frac{m_N}{f_\pi}=10\,, \quad \frac{g_v}{m_v}=1.21\cdot 10^{-2}~\text{MeV}^{-1}\,, \\
		a_2=50\,, \quad a_3 = 5.55\cdot 10^{-3}~\text{MeV}^{-2}\,,\\
		\text{and }a_4 = 6.42\cdot 10^{-5}~\text{MeV}^{-4}\,.
	\end{gathered}
\end{gather}
With the parameters fixed in this way, the nuclear liquid-gas phase transition takes place at the correct values of the chemical potential and saturation density, \mbox{$n_0=0.16\text{ fm}^{-3}$}. Moreover, these parameters were optimized to get realistic values for the compressibility and the surface tension of nuclear droplets. In the next section, we extend the model beyond mean-field level taking into account mesonic fluctuations. 
\end{section}

\begin{section}{Beyond mean field: fluctuations}
A consistent treatment of fluctuations beyond the mean-field approximation can be achieved with the functional renormalization group approach applied to the nucleon-meson model \cite{berges2003quark,berges2000chiral}. We use Wetterich's equation~\cite{wetterich1993exact},
\begin{align}\label{eq:Wetterich}
	\begin{aligned}
		k\,\partial_k\Gamma_k=
		\begin{aligned}
			\vspace{1cm}
			\includegraphics[width=0.08\textwidth]{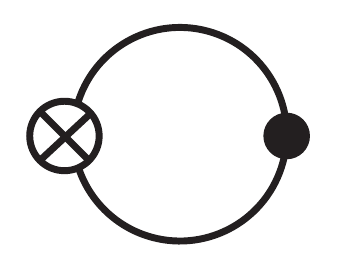}
		\end{aligned} \vspace{-1cm} = \frac 12 \operatorname{Tr}\frac{k\,\partial_k R_k}{\Gamma_k^{(2)}+R_k}\,,
	\end{aligned}
\end{align}
to derive the renormalization group flow of the scale-dependent effective action $\Gamma_k$ under a change of the cutoff scale $k$.
The trace in this flow equation is taken over all bosonic and fermionic degrees of freedom as well as internal indices and involves an integral over space-time or momentum coordinates. The exact inverse propagator $\Gamma_k^{(2)}$ is the second functional derivative of the effective action with respect to the fields. The function $R_k(p)$ regularizes the theory by providing an effective mass for infrared modes. The flow equation (\ref{eq:Wetterich}) connects the bare action, defined at a high-momentum cutoff scale $k=\Lambda$, with the full quantum effective action, $\Gamma_{\text{eff}}$, at $k=0$.
In the actual calculations we apply the leading order of a derivative expansion for which the regulator function can be
optimized~\cite{Litim2000optimisation,litim2001optimised,litim2001mind,pawlowski2007aspects}.
At finite temperature, it is sufficient to regularize the spatial momentum modes, and the appropriate dimensionally reduced regulator function is given by~\cite{litim2006non-perturbative,blaizot2007perturbation}
\begin{gather}
	R_k(\boldsymbol p^2) =(k^2-\boldsymbol p^2)\,\theta(k^2-\boldsymbol p^2)\,.
\label{eq:Litim}
\end{gather}
Since the mass associated with the $\omega$ field is large compared to all relevant energy scales of interest, we continue treating $\omega_0$ as a background field. Nucleons, despite their large mass, can fluctuate around the Fermi surface as particle-hole excitations that are treated properly. The fluctuations of the pion and sigma degrees of freedom are taken into account explicitly.

Previously, the effects of quantum and thermal fluctuations were implicitly parametrized in the effective low-energy potential at $T=0$ and $\mu=\mu_c$. The effective action was then generated by computing the nucleonic loop only. Explicit fluctuation effects of pions and of the sigma field are expected not to be too large. It is reasonable to evaluate their effects as deviations relative to the phenomenological effective potential. Again, only the difference~\eqref{eq:action_diff} with respect to the action at $T=0$ and $\mu=\mu_c$ is relevant. Following~\cite{litim2006non-perturbative} we compute the flow of the difference
\begin{align}
	\bar\Gamma_k(T,\mu)=\Gamma_k(T,\mu)-\Gamma_k(0,\mu_c)
\end{align}
between effective actions at given values of temperature and chemical potential, $\Gamma_k(T,\mu)$, and exactly at the phase transition, $\Gamma_k(0,\mu_c)$. It is given by the flow equation
\begin{align}
	\begin{aligned}
		k\,\partial_k\bar\Gamma_k(T,\mu)&=\phantom{-}\Bigg(
		\begin{aligned}
			\hspace{-.1cm}
			\vspace{1cm}
			\includegraphics[width=0.08\textwidth]{graphics/wetterich_fermion.pdf}
		\end{aligned} \vspace{-1cm}+
		\begin{aligned}
			\vspace{1cm}
			\includegraphics[width=0.08\textwidth]{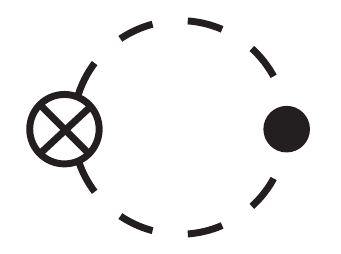}
		\end{aligned} \vspace{-1cm}\Bigg)\Bigg|_{T,\mu}\\
		&\quad-
		\Bigg(
		\begin{aligned}
			\hspace{-.1cm}
			\vspace{1cm}
			\includegraphics[width=0.08\textwidth]{graphics/wetterich_fermion.pdf}
		\end{aligned} \vspace{-1cm}+
		\begin{aligned}
			\vspace{1cm}
			\includegraphics[width=0.08\textwidth]{graphics/wetterich_scalar.pdf}
		\end{aligned} \vspace{-1cm}\Bigg)\Bigg|_{T=0,\mu=\mu_c}\,.
	\end{aligned}
\end{align}
The full circles represent the effects of the nucleons, while the dashed circles are the mesonic loops. The dots indicate full propagators, while the cross-circles stand for the regulator $R_k$. When mesonic loops are ignored, only the nucleons contribute to the flow, and the integration gives their quasiparticle Fermi-gas pressure, as in the mean-field approximation, Eq.~\eqref{eq:UMF}. In leading order of the derivative expansion the effective action takes the form
\begin{align}
	\Gamma_k= \int d^4x\;\left(\frac 12 \partial_\mu\phi^\dagger\,\partial_\mu\phi+ U_k\right)\,,
\end{align}
where $U_k$ is the scale-dependent effective potential. The flow equation simplifies now to an equation for the difference
\begin{align}
	\bar U_k(T,\mu)=U_k(T,\mu)-U_k(0,\mu_c)\,.
\end{align}
For vanishing temperature, the integral extends over all four dimensions with measure $\int \frac{dp_0}{2\pi}\int\frac{d^3p}{(2\pi)^3}$, while for finite temperatures the momentum trace splits into a sum over Matsu\-bara frequencies and a three-dimensional integral over spatial momenta, $T\sum_n\int\frac{d^3p}{(2\pi)^3}$. The integrals and the Matsubara sums can be evaluated explicitly for the spatial Litim regulator (\ref{eq:Litim}). The flow equation for the effective potential $\bar U_k$ becomes
\begin{gather}
	\partial_k\bar U_k(T,\mu)=f(T,\mu)-f(0,\mu_c)
\end{gather}
with
\begin{align}
	\begin{aligned}
		f(T,\mu)&=\frac {k^4}{12\pi^2} \bigg\{\frac{3\big[1+2n_{\text B}(E_\pi)\big]}{E_\pi}+\frac {1+2n_{\text B}(E_\sigma)}{E_\sigma} \\
		&\quad-\frac{8\big[1-n_{\text F}(E_N,\mu_{\text{eff}})-n_{\text F}(E_N,-\mu_{\text{eff}})\big]}{E_N}\bigg\}\,.
	\end{aligned}
\end{align}
Here,
\begin{align}
	\begin{gathered}
		E_\pi^2=k^2+m_\pi^2\,,\quad E_\sigma^2=k^2+m_\sigma^2\,, \quad E_N^2=k^2+g_s^2\sigma^2\,, \\
		m_\pi^2=U_k'(\rho)\,, \quad m_\sigma^2=U_k'(\rho)+2\rho U_k''(\rho)\,, \\
		 \mu_{\text{eff}}=\mu-g_v\,\omega_{0,k}\,,\\
		n_{\text B}(E)=\frac 1{\e^{\beta E}-1}\,,~~\text{ and }~˝\, n_{\text F}(E,\mu)=\frac 1{\e^{\beta(E-\mu)}+1}\,.
	\end{gathered}
\end{align}
In the limit $T \to 0$ the finite-temperature flow equation reduces correctly to the expression obtained at $T=0$ with the $3d$-cutoff function~\cite{braun2003linking,schaefer2005phase}. The prefactors account for the number of degrees of freedom (for nucleons, the number of flavors, $N_f=2$, times a factor of $4$ from the Dirac trace).

In addition to the flow equation for the effective action, the $\omega_0$ field must be computed self-consistently. Therefore, at each momentum scale $k$ we solve the mean-field equation for $\omega_{0,k}$,
\begin{align}
	\frac{\partial U_k}{\partial\omega_{0,k}}=0\,.
\end{align}
The only dependence on $\omega_{0,k}$ appears in the mass term and the fermionic loop. Hence, $\omega_{0,k}$ is given by the solution of the flow equation
\begin{align}
	\partial_k\,\omega_{0,k}=-\frac{2g_v\, k^4}{3\pi^2m_v^2}\,\frac{\partial}{\partial\mu}\left(\frac{n_{\text F}(E_N,\mu_{\text{eff}})+n_{\text F}(E_N,-\mu_{\text{eff}})}{E_N}\right)\,.
\end{align}
In this equation, the effective baryon chemical potential, \mbox{$\mu_{\text{eff}}=\mu-g_v\omega_{0,k}$}, depends also on the field $\omega_{0,k}$, and both $\omega_{0,k}$ and $E_N^2=k^2 + g_s^2 \sigma^2$ depend on the scale $k$. The initial condition for the flow equation is
\begin{align}
	\omega_{0,\Lambda}(\rho) \equiv 0\,.
\end{align}
The ultraviolet scale, $\Lambda$, is a parameter of the model which must be sufficiently large in order to allow for the relevant fluctuation effects and small enough to render the description in terms of the model degrees of freedom realistic; we choose $\Lambda=1.4$ GeV. The flow equation is then solved for a given temperature and chemical potential. The model should be reliably applicable for temperatures up to at least 100~MeV and densities up to about twice the saturation density $n_0$ of nuclear matter. At much higher densities, the field dependence of the Yukawa couplings $g_s$ and $g_v$ can no longer be ignored.

Once fluctuations are taken into account, a readjustment of the potential parameters is required. If the parametrization \eqref{eq:parameterizationMF} is chosen for the potential $U_{\text{MF}}$, the nuclear equilibrium density comes out too low by about ten percent after fluctuations are taken into account. The reason is that the $\mu$ dependence of the thermodynamical potential $U$ is more involved due to the influence of the mesonic fluctuations. It is necessary to readjust the parameters in such a way that the nuclear physics constraints are reproduced in the presence of fluctuations. The parameters of the potential used in the following are:
\begin{align}
	\begin{gathered}
		g_s=10\,,\quad\frac{g_v}{m_v}=1.02\cdot 10^{-2}~\text{MeV}^{-1}\,, \\
		a_2=65.9\,,\quad a_3 = 5.55\cdot 10^{-3}~\text{MeV}^{-2}\,,\\
		\text{and }a_4 = 8.38\cdot 10^{-5}~\text{MeV}^{-4}\,.
	\end{gathered}
\end{align}
The resulting nuclear matter quantities are listed in Table \ref{tab:nucl_phys}. The mass of the $\sigma$ boson (not to be confused with the position of the complex pole at $\sqrt{s}\sim (500 -i\,300)$ MeV in the $I=0$ \textit s-wave $\pi\pi$ \textit T matrix \cite{caprini2006mass,yndurain2007experimental}) becomes $m_\sigma \simeq 770$ MeV with inclusion of mesonic fluctuations. Not surprisingly, it is significantly larger than the sigma mass used previously at the level of mean-field phenomenology ($m_\sigma \simeq 670$ MeV). The compression modulus of nuclear matter is
\begin{align}
	K=9n\left(\frac{dn}{d\mu}\right)^{-1}\,,
\end{align}
where $n$ is the baryon density. The nuclear surface tension is computed in the model as \cite{berges2003quark}
\begin{align}
	\Sigma=\int_{\sigma_0}^{f_\pi}d\sigma\;\sqrt{2U}\,,
\end{align}
where $U=U_{k=0}$ is the effective potential and $\sigma_0$ is the value of the sigma field at normal nuclear matter density.
\begin{table}
	\begin{tabular}{lll}
		& Model & Empirical \\
		\hline
		Nucl. saturation density $n_0$ & $0.156$ fm${}^{-3}$ & $0.16$ fm${}^{-3}$ \\
		Binding energy $B$& 16.3 MeV & 16 MeV \\
		Surface tension $\Sigma$ & 1.11 MeV/fm${}^2$ & 1.1 MeV/fm${}^2$ \\
		Compression modulus $K$ & 330 MeV  & $240\pm 30$ MeV 
	\end{tabular}
	\caption{Comparison of the fitted model parameters with empirical data.\label{tab:nucl_phys}}
\end{table}
\end{section}

\begin{section}{Results and discussion}
\begin{subsection}{Liquid-gas transition}
\begin{figure}
	\centering
	\includegraphics[width=0.4\textwidth]{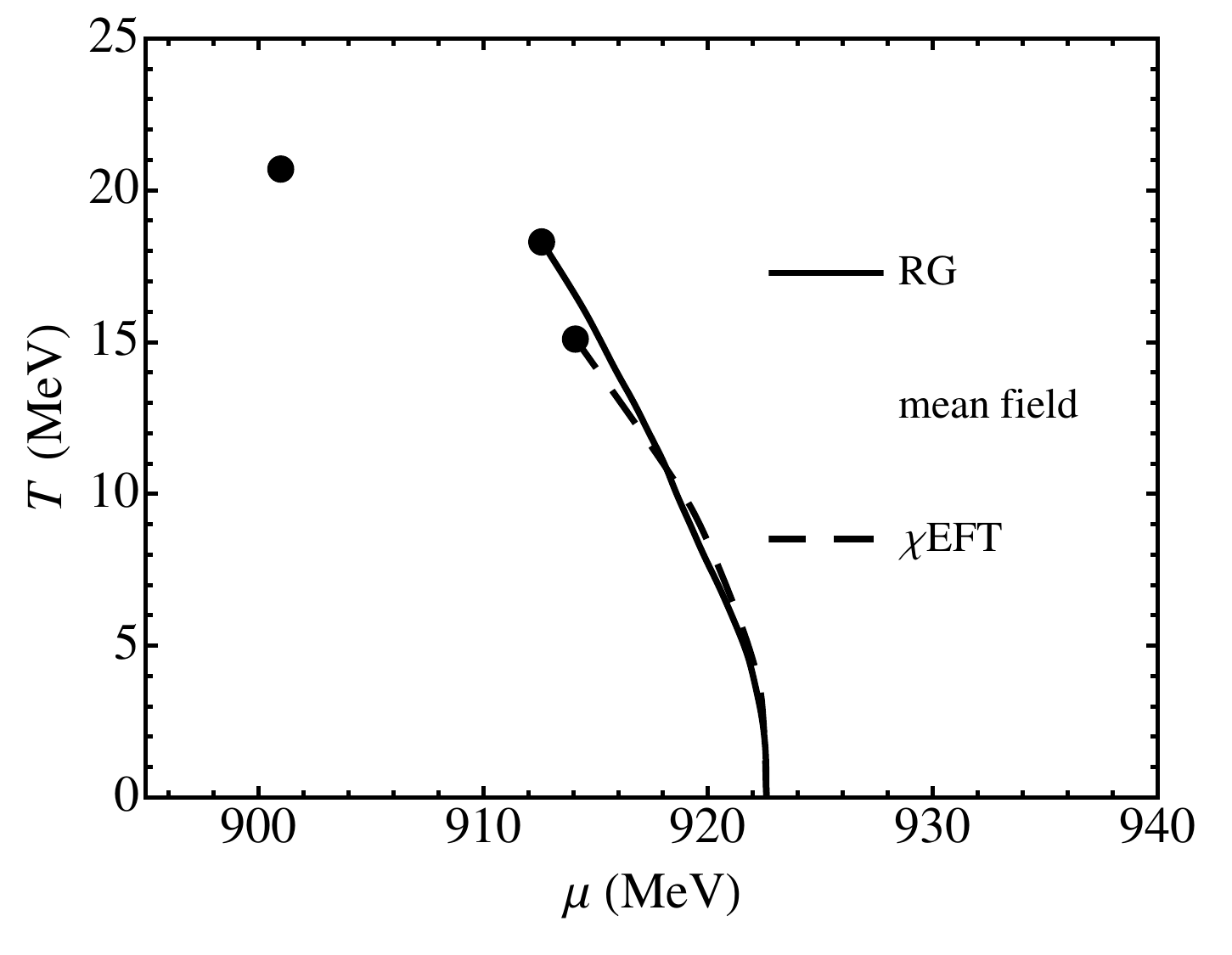}
	\vspace{-0.2cm}
	\caption{Liquid-gas phase transition. Dotted curve: mean-field result of the chiral meson-nucleon model. Solid curve: FRG calculation including mesonic fluctuations. Dashed curve: in-medium chiral effective field theory calculation of Refs.~\cite{fiorilla2012chiral,fiorilla2012nuclear}.\label{fig:liquidgasphase}}
\end{figure}

The flow equations are now solved using the grid method proposed in~\cite{Adams1995Solving}. The resulting system of coupled differential equations for the effective potential and its derivatives is solved numerically. The thermodynamic potential is derived by minimizing the effective potential as a function of the constant fields $\sigma$ and $\omega_0$.

As mentioned, the present model is supposed to be reliable in the regime around normal nuclear matter including the liquid-gas phase transition, and up to about twice the equilibrium density, $n_0$. We recall that the average distance between nucleons at normal nuclear density, $d\simeq 1.9$~fm, scales with the third root of the density and reduces to about 1.5~fm at $2n_0$. In chiral pictures of the nucleon with a compact baryonic core surrounded by a pionic cloud, the cores are still well separated at such densities.

In the following, we shall compare results from our FRG approach with calculations in chiral effective field theory (\cite{Holt2013Nuclear} and references therein). We proceed by studying, first, the nuclear liquid-gas phase transition and the equation of state, then the density and temperature dependence of the chiral condensate, the order parameter of spontaneously broken chiral symmetry. 

Consider the nuclear liquid-gas phase transition in the \mbox{$T$-$\mu$} diagram. Figure~\ref{fig:liquidgasphase} shows the first-order transition boundary. The bending of this curve is understood from a Clausius-Clapeyron type relation. Along the phase transition, the total differentials of the effective potential must agree in the liquid and in the gaseous phases:
\begin{align}
	\frac{\partial U_{\text{liquid}}}{\partial\mu}d\mu+\frac{\partial U_{\text{liquid}}}{\partial T}dT=\frac{\partial U_{\text{gas}}}{\partial\mu}d\mu+\frac{\partial U_{\text{gas}}}{\partial T}dT\,.
\end{align}
The slope of the transition line is therefore given by the ratio of differences between baryon number densitites, $n_{\text{liquid}}-n_{\text{gas}}$, and entropy densities, $s_{\text{liquid}}-s_{\text{gas}}$, as follows
\begin{align}
	\frac{dT}{d\mu}=-\frac{n_{\text{liquid}}-n_{\text{gas}}}{s_{\text{liquid}}-s_{\text{gas}}}\,.
\end{align}
By comparing the mean-field result of the chiral nucleon-meson model~\cite{floerchinger2012chemical} (dotted curve in Fig.~\ref{fig:liquidgasphase}) with the FRG calculation (solid curve) it is apparent that fluctuations beyond the mean field bend the phase-transition boundary toward higher chemical potentials as the temperature increases. In the mean-field approximation without mesonic fluctuations, the entropy is entirely determined by the nucleons. For small temperatures and a chemical potential below $\mu_c$, no Fermi sea of nucleons exists. For $\mu >\mu_c$, the Fermi sphere is filled and particle-hole excitations around the Fermi surface contribute to the entropy. Therefore the entropy is larger in the liquid phase than in the gas phase and since $n_{\text{liquid}}>n_{\text{gas}}$, the slope of the $T-\mu$ phase boundary is negative, $\frac{dT}{d\mu}<0$, as observed.

The curvature of the boundary line is in good agreement with the $\chi$EFT results of Refs.~\cite{fiorilla2012chiral,fiorilla2012nuclear}. This is a nontrivial observation since the two approaches, FRG versus $\chi$EFT, differ significantly in their treatment of fluctuations associated with the pion field and its thermodynamics. The $\chi$EFT calculations are based on a perturbative expansion of the free-energy density up to three-loop order, including all one- and two-pion exchange processes in the medium together with three-body forces and effects from $\Delta$-isobar excitations. The FRG approach involves a nonperturbative resummation of pion and nucleon loops but relegates many other effects to the parametrization of the effective potential $U$. The FRG critical point of the liquid-gas transition lies at slightly higher temperature than the one in three-loop $\chi$EFT: one finds a critical temperature $T_c=18.3\text{ MeV}$, compared to the $\chi$EFT result, $T_c=15.1\text{ MeV}$~\cite{fiorilla2012chiral,fiorilla2012nuclear}. This is consistent with estimates from multifragmentation and fission data which place the critical temperature at $T_c\gtrsim 16\text{ MeV}$~\cite{Karnaukhov2008Critical}. 

\begin{figure}
	\centering
	\includegraphics[width=0.4\textwidth]{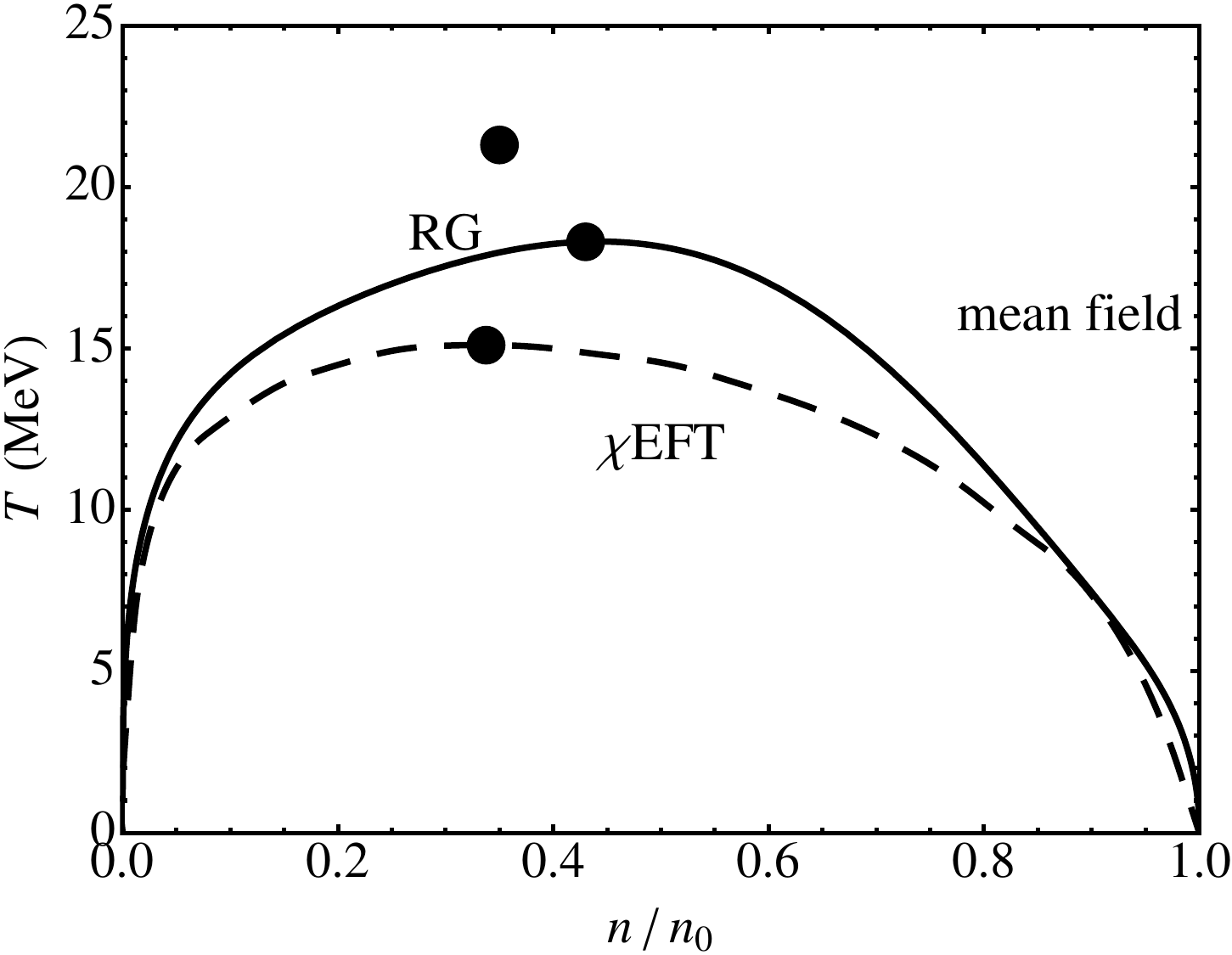}
	\vspace{-0.2cm}
	\caption{Coexistence region of the liquid and fluid phase in mean-field approximation and with fluctuations compared to $\chi$EFT~\cite{fiorilla2012chiral,fiorilla2012nuclear}. The black dots label the respective critical endpoint.\label{fig:Trho}}
\end{figure}

The liquid-gas coexistence region plotted in the temperature-density plane is shown in Fig.~\ref{fig:Trho}. It features, as in Fig.~\ref{fig:liquidgasphase}, a calculation in mean-field approximation, the result with fluctuations treated in the FRG framework, and the $\chi$EFT result.

\begin{figure}
	\centering
	\includegraphics[width=0.4\textwidth]{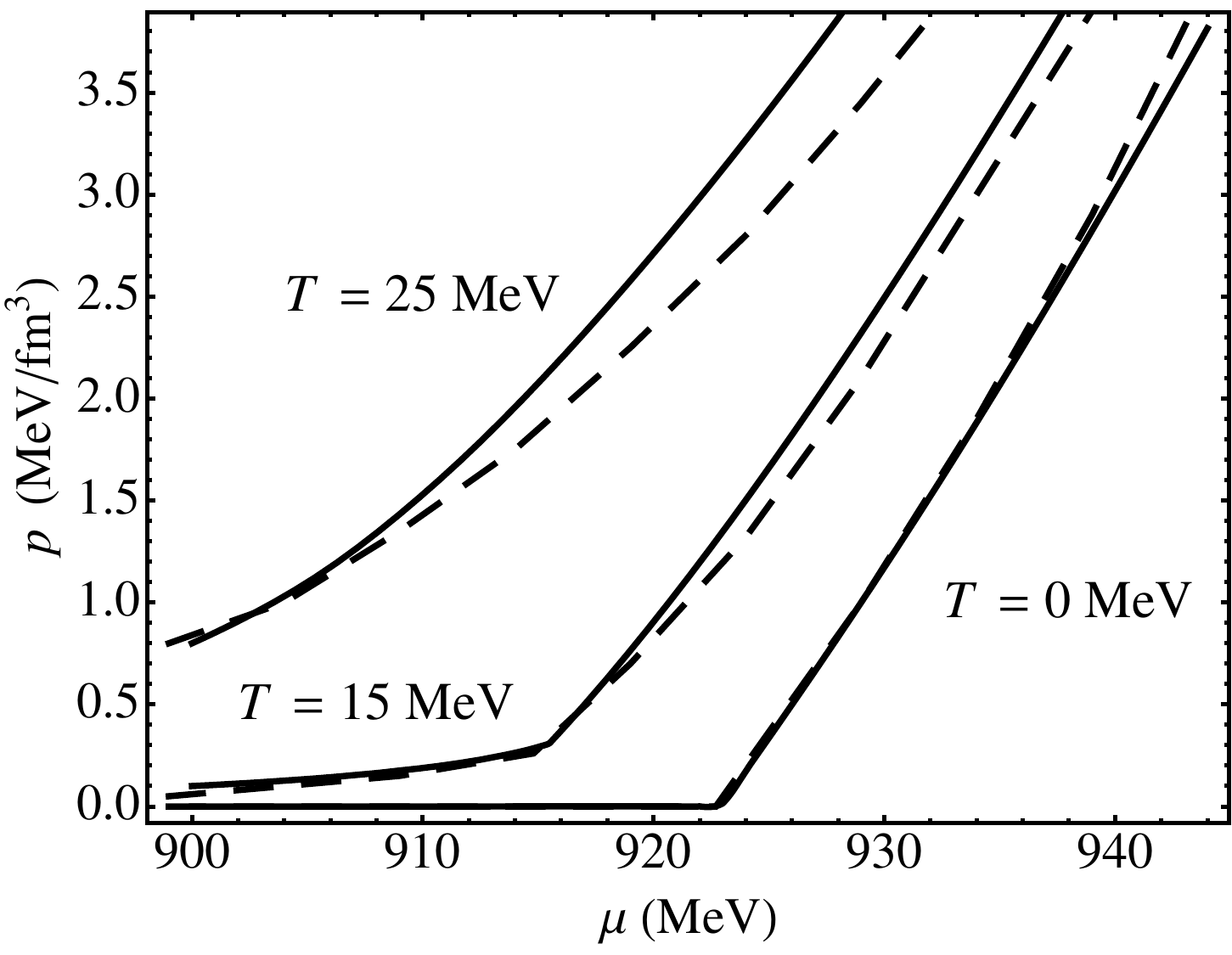}
	\vspace{-0.2cm}
	\caption{Pressure as a function of baryon chemical potential in the FRG model calculation (full line) compared with $\chi$EFT~\cite{fiorilla2012chiral,fiorilla2012nuclear} (dashed line) for three different temperatures.\label{fig:p_mu}}
\end{figure}

A comparison between the pressure $P(\mu)$ resulting from the model with inclusion of RG effects and from $\chi$EFT is shown in Fig.~\ref{fig:p_mu}. Since the effective potential is adjusted to reproduce nuclear observables at $\mu=\mu_c$ and $T=0$, the equations of state agree very well in both approaches around this point. In particular, the slope of $P(\mu)$ at $\mu_c$ is related to the compressibility which is reasonably consistent with the empirical compression modulus in both FRG and $\chi$EFT calculations. The equations of state match also for larger chemical potentials at $T=0$. As the temperature increases some deviations between the FRG and $\chi$EFT equations of state appear, although they remain small for temperatures up to 15--20~MeV. These features reflect the similarity of the first-order transition lines in the phase diagram, with the exception of the small relative displacement in the position of the critical endpoint. Given the different treatments of the pionic physics in the FRG and $\chi$EFT approaches, the close similarity of these results is once again remarkable.

Next, consider the chiral condensate, $\left\langle\bar qq\right\rangle$, as a function of temperature and baryon density (or chemical potential). In the chiral nucleon-meson model this condensate is proportional to the expectation value of the $\sigma$ field. Quite generally, the Hellmann-Feynman theorem in combination with the Gell-Mann--Oakes--Renner relation gives the in-medium chiral condensate in the form~\cite{fiorilla2012chiral,fiorilla2012nuclear}
\begin{align}
	\frac{\left\langle\bar qq\right\rangle(n,T)}{\left\langle0|\bar qq|0\right\rangle}
=1-{\partial{\cal F}(n,T)\over f_\pi^2\,\partial m_\pi^2}\,,
\end{align}
where $\mathcal F = \mathcal E - Ts$ is the free-energy density, $\mathcal F=n\bar{F}$ with $\bar{F}$ the free energy per particle. $\langle 0|\bar qq|0 \rangle$ is the chiral condensate in vacuum. The pion-mass dependence of $\bar{F}$ is a quantity systematically accessible in $\chi$EFT since this dependence is explicitly given in terms of the pion propagators present in the in-medium loop diagrams.

\begin{figure}
	\includegraphics[width=0.4\textwidth]{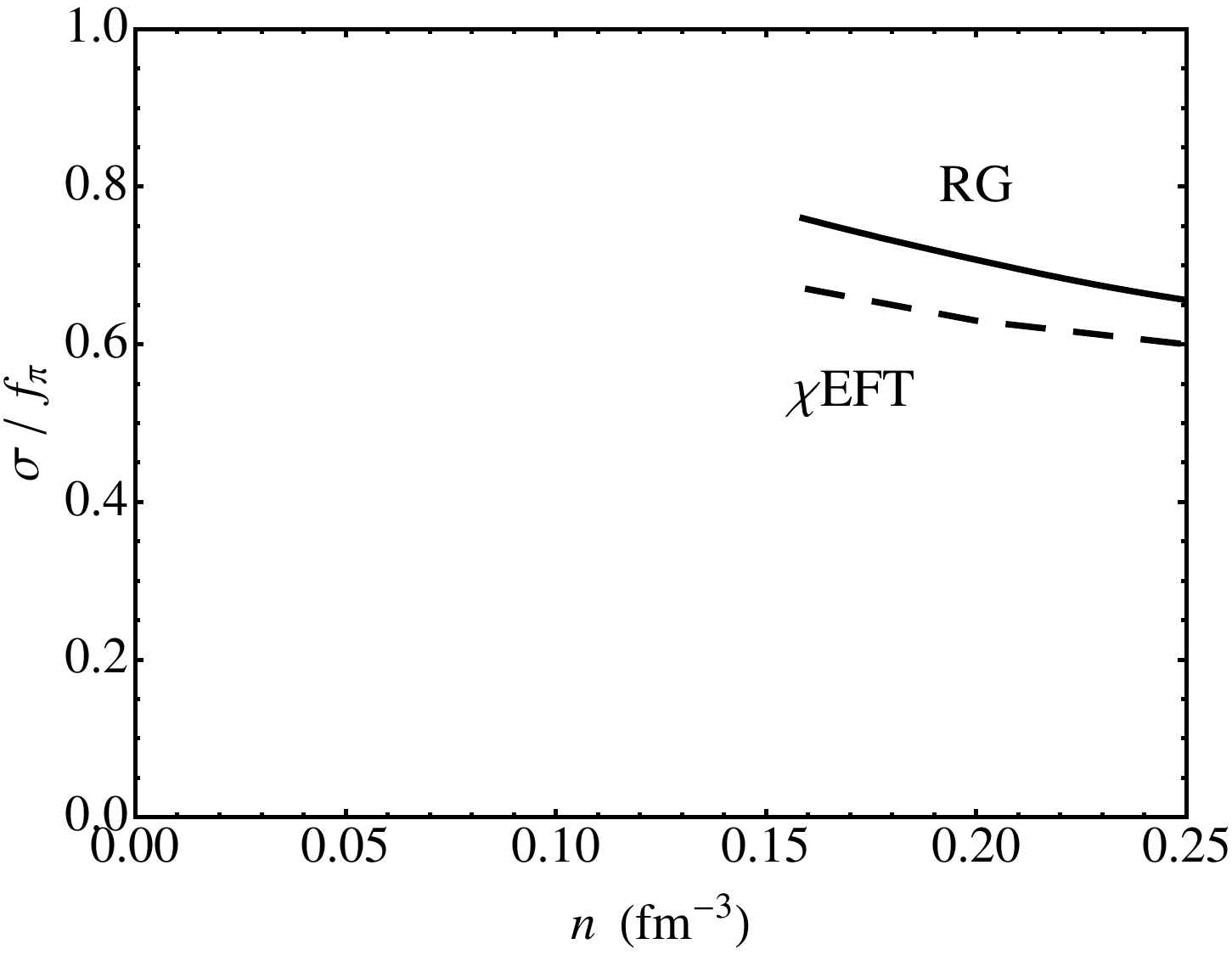}
	\vspace{-0.2cm}
	\caption{Chiral order parameter versus density at $T=0$. The dotted lines are obtained by applying a  Maxwell construction along the nuclear liquid-gas coexistence region. The RG result (solid curve) is shown in comparison with $\chi$EFT~\cite{fiorilla2012chiral,fiorilla2012nuclear} (dashed).\label{fig:SigmaRho}}
\end{figure}
\begin{figure}
	\centering
	\includegraphics[width=0.4\textwidth]{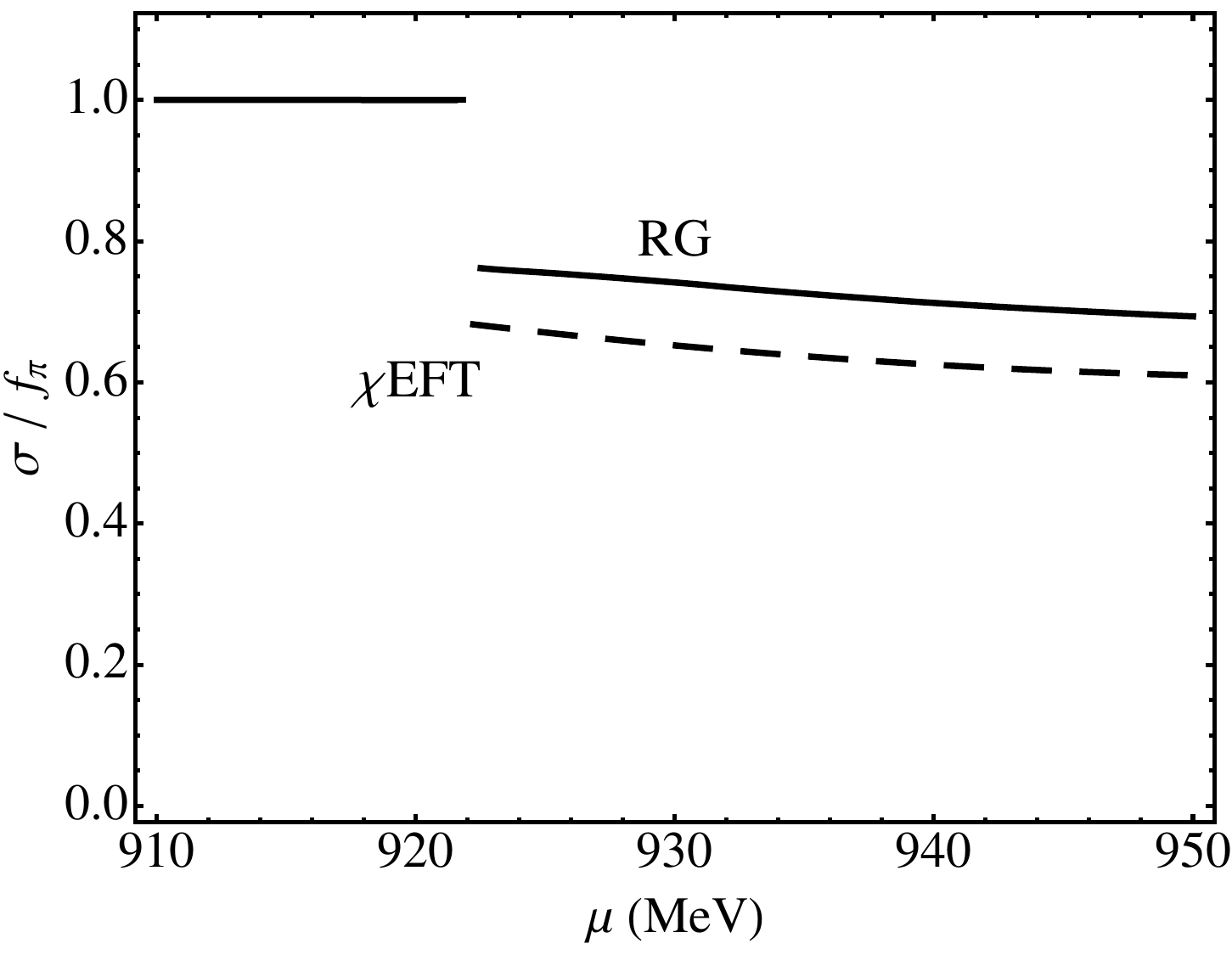}
	\vspace{-0.2cm}
	\caption{Chiral order parameter at $T=0$ as a function of baryon chemical potential, calculated in the chiral nucleon-meson model (solid curve) with inclusion of fluctuations using the FRG approach. The $\chi$EFT result (dashed curve) is taken from~\cite{fiorilla2012chiral,fiorilla2012nuclear}.\label{fig:SigmaMu}}
\end{figure}

Figures \ref{fig:SigmaRho} and \ref{fig:SigmaMu} show the chiral condensate at zero temperature as functions of the baryon chemical potential $\mu$ and density $n$, plotted as the ratio of $\sigma$ versus its vacuum value $\sigma_0$. The density dependence of the condensate at $T=0$ displayed in Fig.~\ref{fig:SigmaRho} shows (dotted) the behavior in the presence of the liquid-gas coexistence region up to the equilibrium density of normal nuclear matter. At higher densities, correlations and fluctuations beyond mean field tend to stabilize the chiral condensate against restoration of chiral symmetry in its Wigner-Weyl realization, at least up to about twice $n_0$, the conservative estimate for the range of applicability of the present investigation. The presentation of the chiral condensate as a function of baryon chemical potential (Fig.~\ref{fig:SigmaMu}) is particularly instructive as it demonstrates the impact of the first-order liquid-gas transition on an order parameter of completely different origin, manifest in the discontinuity at $\mu=\mu_c=923\text{ MeV}$. At larger baryon chemical potential there is no tendency towards rapid chiral symmetry restoration. Pionic fluctuations delay the dropping of the condensate. In this respect the FRG treatment shows an even more pronounced effect than the three-loop $\chi$EFT calculations, though it is again remarkable how close the (nonperturbative) FRG results and the (perturbative) $\chi$EFT results turn out to be.
\end{subsection}

\begin{subsection}{Chemical freeze-out and chiral phase transition}
\begin{figure}
	\centering
	\includegraphics[width=0.4\textwidth]{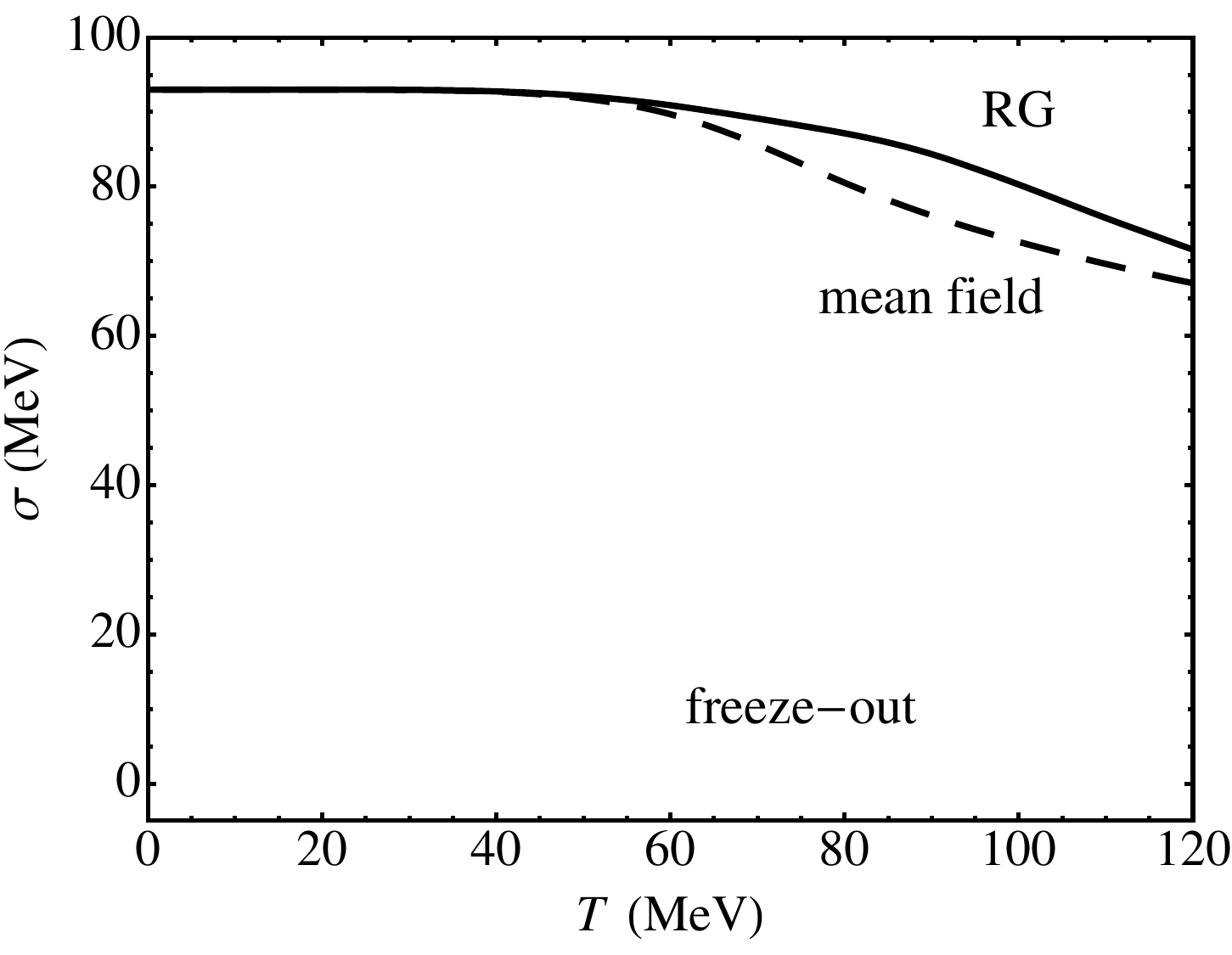}
	\vspace{-0.2cm}
	\caption{$T$ dependence of the chiral condensate at $\mu=760$~MeV. The curves at the mean-field level and with fluctuations included are compared. The experimental freeze-out point is at $T=56^{+9.6}_{-2.0}$~MeV for $\mu=760^{+23}_{-23}$~MeV~\cite{Andronic2009Thermal}.\label{fig:mu760}}
\end{figure}

Abundances of hadronic species produced in heavy-ion collisions are well described in a hadronic resonance gas picture. Using a statistical model a chemical freeze-out boundary curve in the $(T,\mu)$ diagram has been extracted~\cite{Andronic2009Thermal}. For small baryon chemical potentials the freeze-out temperature turns out to be very close to the transition temperature of the chiral crossover as inferred from lattice QCD computations. If such a correspondence between chemical freeze-out and chiral transition remained valid also for large chemical potentials, one would be tempted to conclude that the chiral phase transition leaks well into the nuclear physics terrain that is properly described by the present chiral nucleon-meson model.  It is therefore of interest to explore whether the model as it stands would support or disprove such an interpretation.

A partial answer has already been given in Ref.~\cite{floerchinger2012chemical}. Their mean-field analysis  shows no decreasing chiral condensate near freeze-out at large chemical potentials. Here we repeat and extend this computation, now with the effects from pion loops included. As a typical example, the $\sigma$ field representing the chiral condensate is plotted as a function of temperature for a fixed chemical potential $\mu=760$~MeV in Fig.~\ref{fig:mu760}. At this value of $\mu$ the freeze-out point derived from the statistical model analysis is located at a temperature $T=56^{+9.6}_{-2.0}$~MeV. If there were a chiral phase transition nearby, the condensate would change significantly and drop rapidly to a small value. This is not seen in Fig.~\ref{fig:mu760} where the sigma field is plotted both at the mean-field level and with the fluctuations taken into account using the FRG. One observes that the magnitude of the chiral condensate is still large up to temperatures around 100~MeV and chiral symmetry remains spontaneously broken. Chemical freeze-out and chiral restoration are not connected or intertwined at baryon chemical potentials characteristic of the nuclear physics region and beyond.

In Fig.~\ref{fig:condensate}, the contours of the normalized condensate, $\sigma/f_\pi$, are plotted for chemical potentials in the range \mbox{$700\text{ MeV}\le\mu\le1\text{ GeV}$}. We see that the condensate stays above $2/3$ of its vacuum value throughout this region. We therefore conclude that chiral symmetry is not restored and there is no critical endpoint within the region \mbox{$700\text{ MeV}\le\mu\le1\text{ GeV}$} and for temperatures \mbox{$T\le 100\text{ MeV}$}.

It should, of course, be pointed out that the chiral phase transition or the crossover itself cannot be reliably addressed in our model. The effective potential has been adjusted at the liquid-gas phase transition in a Taylor expansion around $\sigma=f_\pi$. It is therefore predictive only for values of $\sigma$ not too far from $f_\pi$, whereas $\sigma$ changes rapidly in the vicinity of the phase transition or crossover. It is nonetheless instructive to extrapolate and examine where the phase transition actually takes place in the model. In the mean-field approximation, the condensate is seen to jump discontinuously to zero already at a chemical potential of $\mu=970$~MeV which translates to a density of about $1.75$ times saturation density. This restricts the applicability of the mean-field version to a relatively narrow range around 
normal nuclear densities and the liquid-gas transition. Once fluctuations are properly treated using the FRG approach, the chiral condensate remains finite up to a chemical potential $\mu=1.15$~GeV, or densities beyond \mbox{$2.5$ times} nuclear saturation density. While at such large values of the chemical potential, the field dependence of the Yukawa couplings should already be taken into account, the fact that fluctuations tend to stabilize the hadronic phase of spontaneously broken chiral symmetry up to quite high baryon densities emerges as a robust result.
\end{subsection}

\begin{figure}
	\centering
	\includegraphics[width=0.4\textwidth]{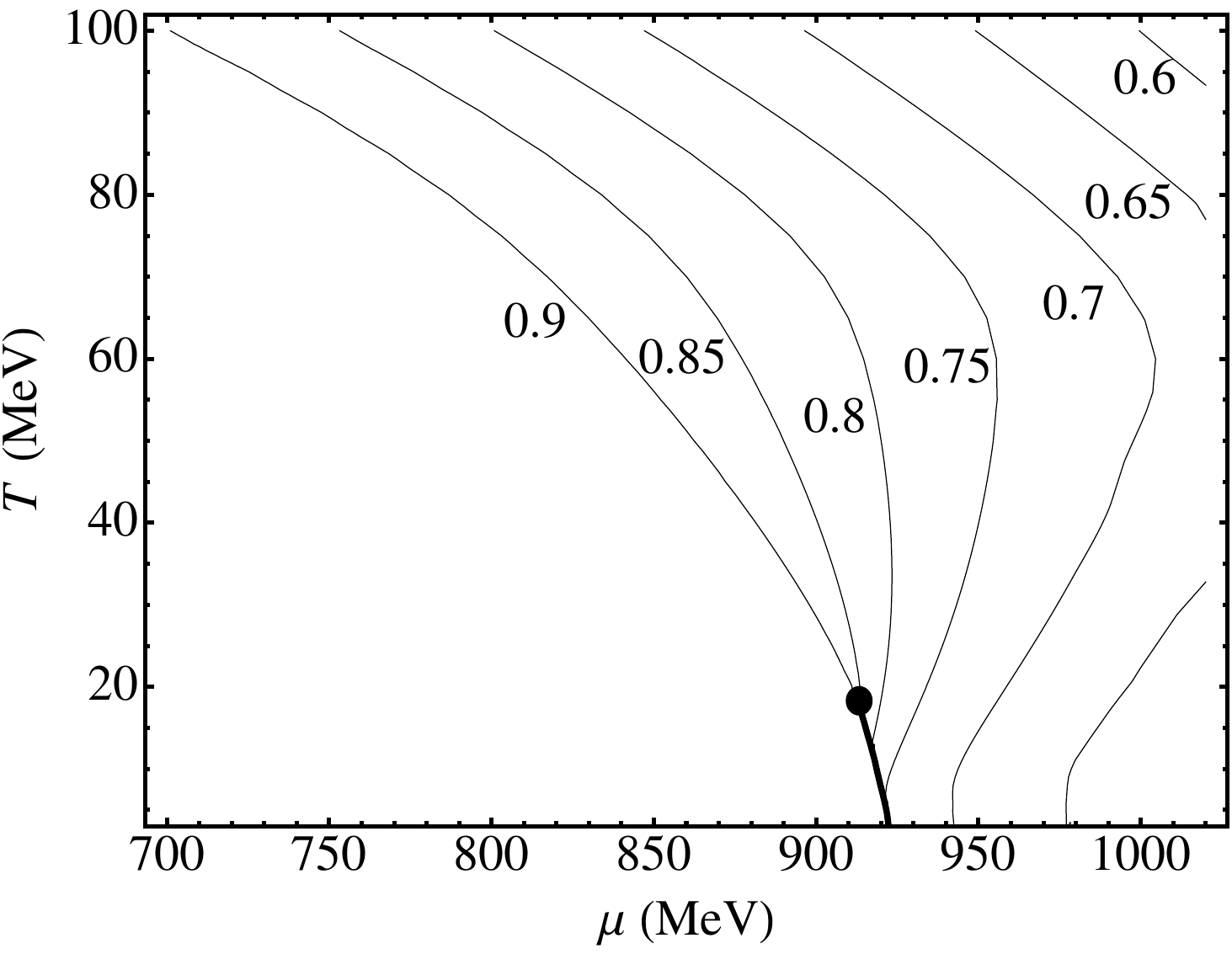}
	\vspace{-0.2cm}
	\caption{Contour plots of $\sigma/f_\pi$. Within the region of applicability ($\mu \lesssim 1$ GeV, $T\lesssim 100$ MeV) of the chiral nucleon-meson model, the chiral condensate is nonzero and chiral symmetry is not restored. \label{fig:condensate}}
\end{figure}

\begin{subsection}{Fluctuation effects near the critical endpoint of the liquid-gas transition}
\begin{figure}
	\centering
	\includegraphics[width=0.4\textwidth]{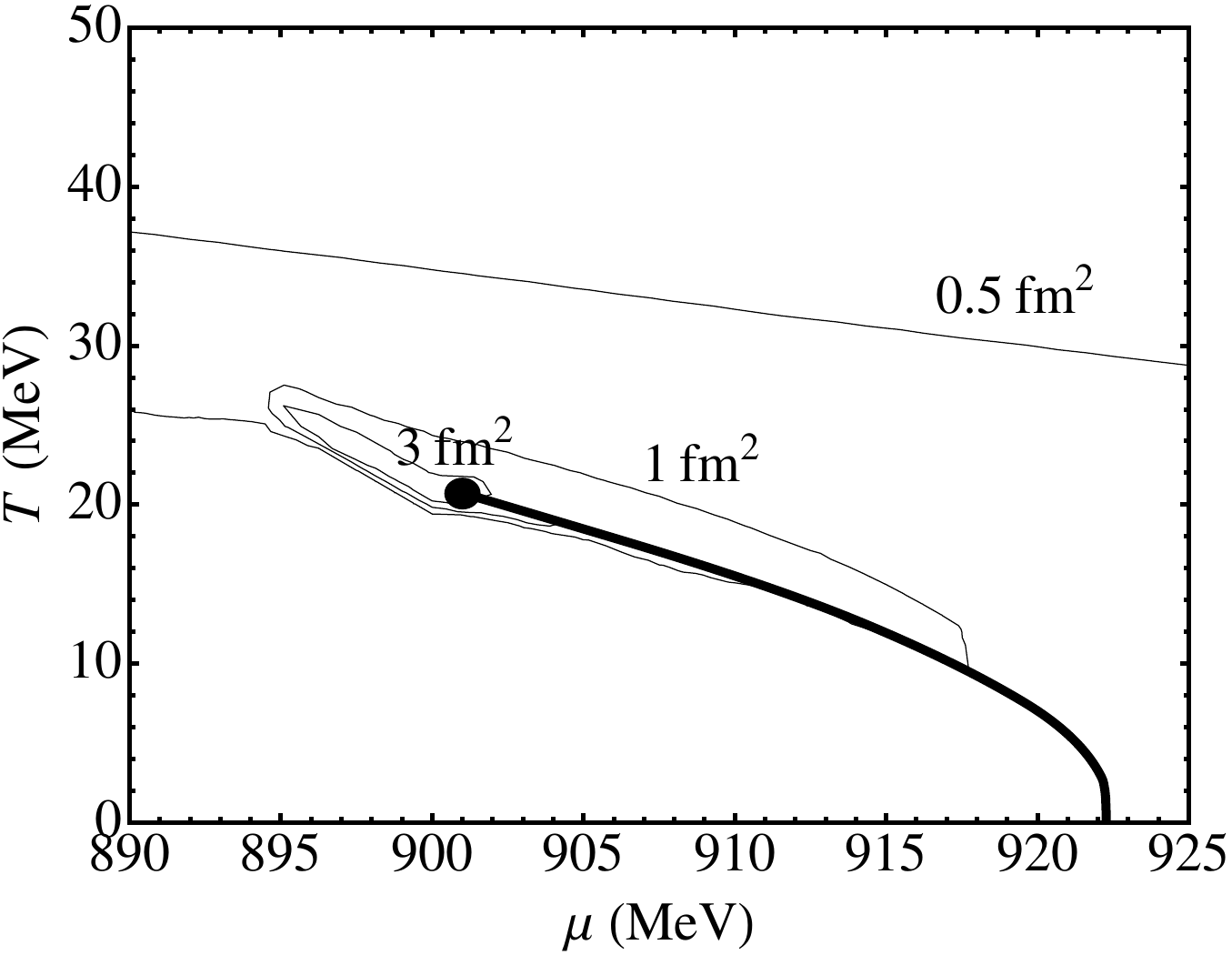}
	\vspace{-0.2cm}
	\caption{Contour plot of the baryon number susceptibility $\chi_n(\mu, T) =\frac{\partial n}{\partial\mu}$ from the mean-field calculation in the $T$-$\mu$ plane around the critical endpoint of the liquid-gas phase transition. \label{fig:flucdensityMF}}
\end{figure}
\begin{figure}
	\centering
	\includegraphics[width=0.4\textwidth,clip=true]{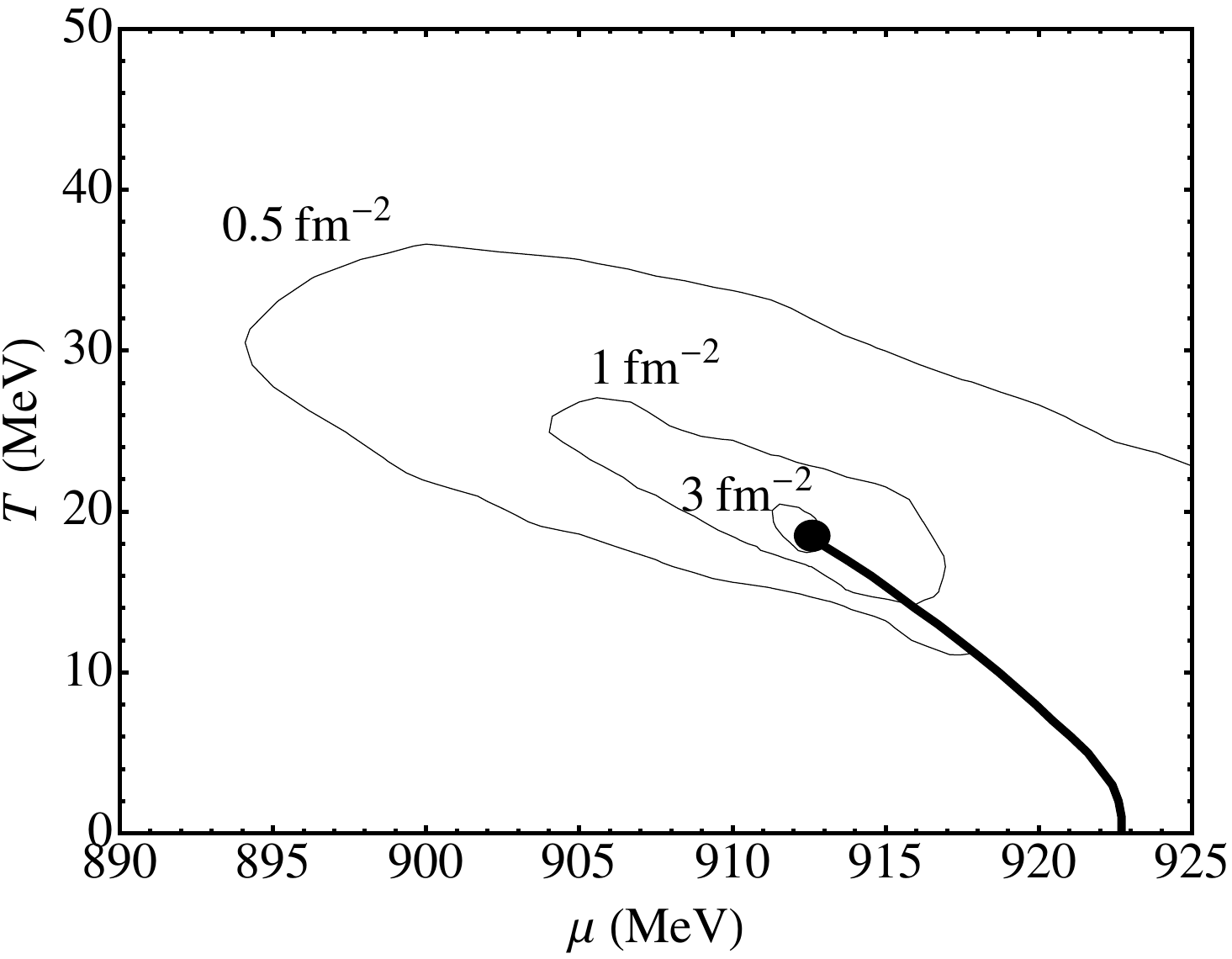}
	\vspace{-0.2cm}
	\caption{Same plot as in Fig.~\ref{fig:flucdensityMF}, but now with fluctuations taken into account.\label{fig:flucdensityRG}}
\end{figure}
The fluctuation effects included in the present FRG calculation are also important for the description of critical behavior in the vicinity of the endpoint of the first-order liquid-gas transition. As already discussed in Ref.~\cite{floerchinger2012chemical} for the present model, a mean-field calculation cannot be expected to be reliable close to the phase transition.

To assess the magnitude of these fluctuations near the critical point, we compare in the following results for the baryon number and chiral susceptibilities from the FRG calculation to those from the mean-field approach. A technically similar calculation~\cite{schaefer2007susceptibilities} for the critical region in a quark-meson model found only a relatively narrow region around the critical endpoint (in this case of the chiral phase transition) in which fluctuations dominate. Compared to the mean-field calculation, the critical region in those FRG results was much compressed. While the calculations performed with the quark-meson model were focused on quark-number susceptibilities, the results guide our expectations also for the present model. In the PQM study~\cite{skokov2010vacuum}, a smoothing of the observables around the chiral crossover line appeared once fluctuations were included.

The liquid-gas phase transition is characterized by a discontinuity in baryon density as a function of baryon chemical potential. The corresponding baryon number susceptibility,
\begin{align}
	\chi_n=-\frac{\partial^2U}{\partial\mu^2}=\frac{\partial n}{\partial\mu}\,,
\end{align}
is discontinuous at the first-order liquid-gas phase transition and diverges at the second-order critical endpoint.
In Figs.~\ref{fig:flucdensityMF} and \ref{fig:flucdensityRG}, contour lines of $\chi_n$ are plotted in the $T$-$\mu$ diagram. In the nucleon-meson model there is an extended region above the critical endpoint where the susceptibility in the mean-field calculation remains large. This region is elongated along an extrapolation of the first-order line beyond the critical endpoint. In contrast, the fluctuation-dominated region in the FRG calculation is more concentrated around the critical endpoint and smaller in extent, in particular in the direction of the chemical potential.

The baryon density is the order parameter of the liquid-gas transition. But this first-order phase transition leaves its traces also in the chiral order parameter, $\sigma$, in the form of a discontinuous jump. It is interesting to examine the corresponding susceptibility, $\chi_\sigma=m_\sigma^{-2}$. In Figs.~\ref{fig:flucMF} and \ref{fig:flucRG}, contour lines of the chiral susceptibility are shown in the $T$-$\mu$ plane. To facilitate a comparison, the susceptibilities are normalized to their respective vacuum expectation values according to $\chi_\sigma(\mu, T) \times m_{\sigma, \mathrm{vac}}^2$. The results are qualitatively similar to those for the density fluctuations. In the mean field analysis, the region of large susceptibilities extends to smaller chemical potentials away from the critical endpoint, whereas these regions are more centered around the critical endpoint once fluctuations are taken into account.

\begin{figure}
	\centering
	\includegraphics[width=0.4\textwidth]{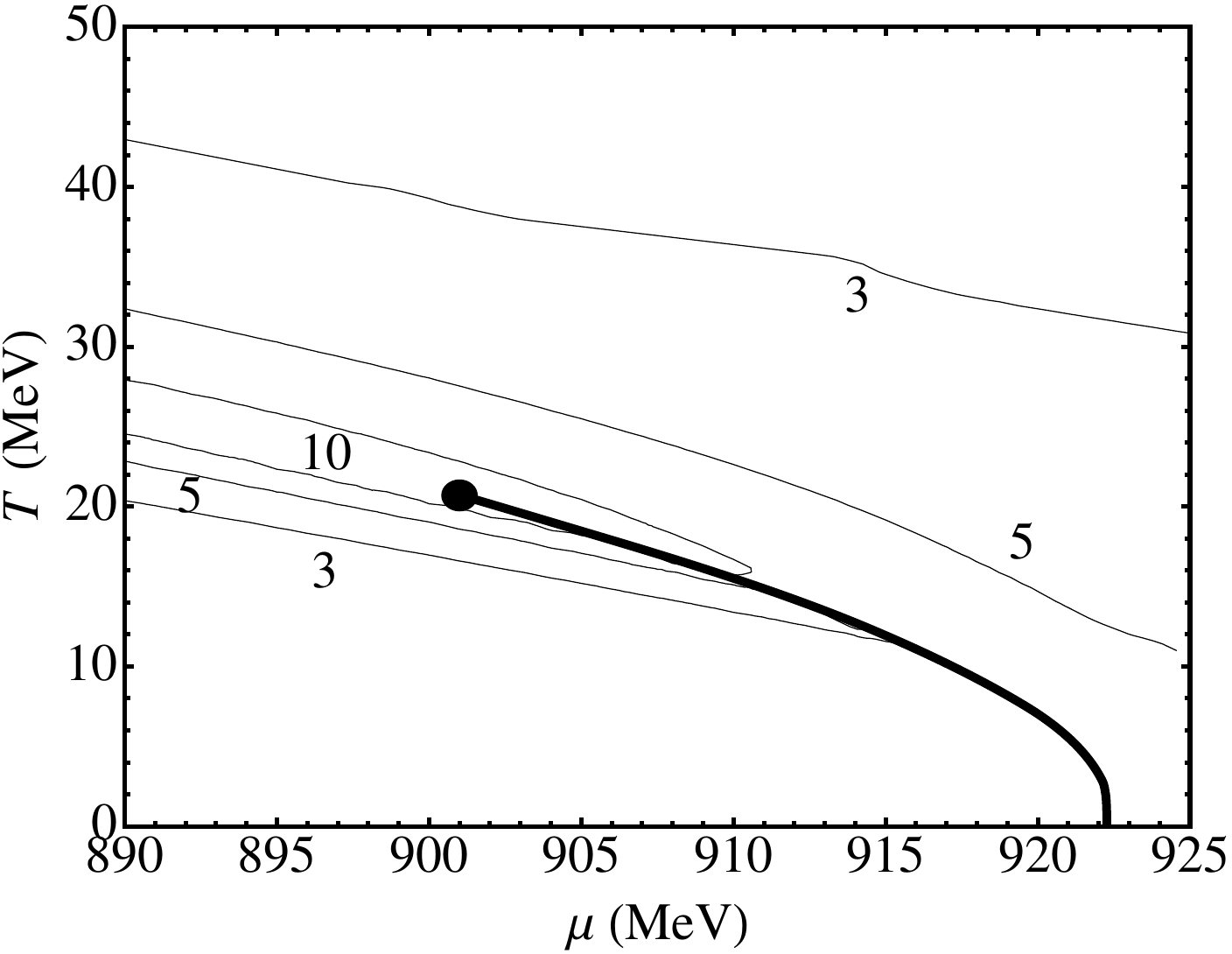}
	\vspace{-0.2cm}
	\caption{Contour plot of the normalized chiral susceptibility $\chi_\sigma(\mu, T) \cdot m_{\sigma, \mathrm{vac}}^2$ from the mean-field calculation in the $T$-$\mu$ plane around the critical endpoint of the liquid-gas phase transition. \label{fig:flucMF}}
\end{figure}
\begin{figure}
	\centering
	\includegraphics[width=0.4\textwidth,clip=true]{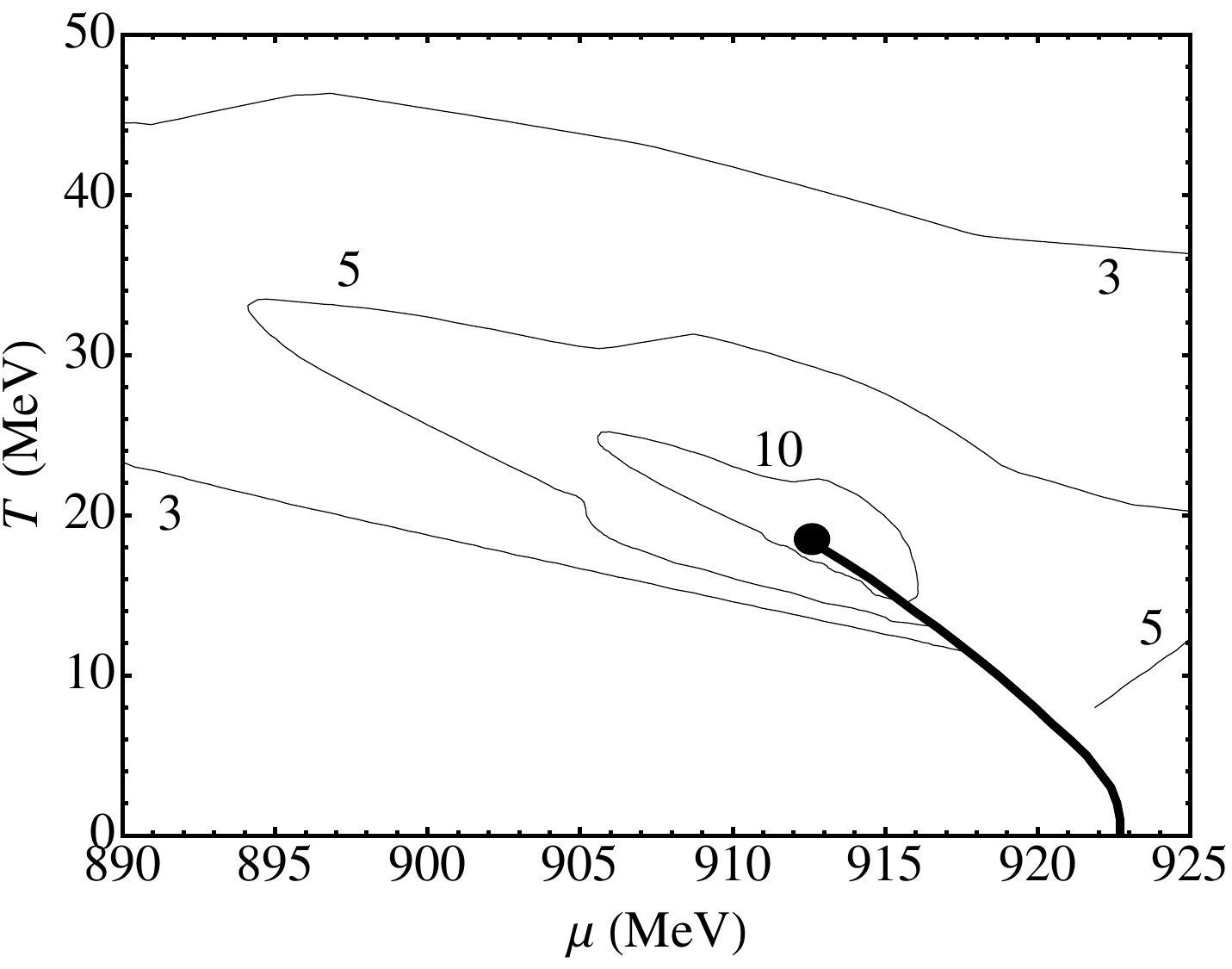}
	\vspace{-0.2cm}
	\caption{Same plot as in Fig.~\ref{fig:flucMF}, but now with fluctuations taken into account.\label{fig:flucRG}}
\end{figure}
\end{subsection}
\end{section}

\begin{section}{Summary}
The present work provides an extension of the chiral nucleon-meson model including effects of fluctuations treated within the framework of the functional renormalization group. The few parameters of the model are fitted to reproduce physical observables around the nuclear liquid-gas phase transition at $\mu=\mu_\mathrm{c}$ and $T=0$. A vector interaction is included and the expectation value of the corresponding background field is determined self-consistently. 

The liquid-gas phase transition determined in this calculation is compared with results from in-medium chiral effective field theory. The first-order phase transition lines at low temperatures agree well, while a slightly larger temperature for the critical endpoint is found in our present treatment, in good agreement with the empirical range of critical temperatures. The chiral condensate decreases more slowly as a function of baryon chemical potential compared to mean-field results. This is, again, in agreement with $\chi$EFT where higher-loop effects, the $\Delta$-isobar resonance and three-body forces lead to a similar behavior. The equations of state resulting from the two approaches agree for the first-order liquid-gas phase transition at $T=0$ where the model effective potential is fitted to physical observables. This agreement holds also for larger values of the chemical potential as well as for temperatures up to 15--20~MeV.

In comparison with mean-field results, we find that fluctuations exert a stabilizing effect on the phase diagram: fluctuation effects are more focused around the critical endpoint of the liquid-gas phase transition, whereas the mean-field results show large chiral and baryon number susceptibilities along the extrapolation of the first-order line. These findings in the chiral nucleon-meson model have a qualitative correspondence in observations made in studies of the chiral phase transition using the quark-meson model.  

We have found no hints of a first-order chiral transition and its critical endpoint in the regime of applicability of the model. The chiral order parameter decreases slowly, as in $\chi$EFT, and the first-order chiral phase transition is delayed to larger values of the baryon chemical potential. The order parameter remains large up to temperatures of 100~MeV and baryon chemical potentials of at least about 1~GeV. The present results demonstrate that it is crucial to include nucleons and their correlations in order to correctly reproduce the equation of state in the hadronic phase of QCD around the nuclear liquid-gas transition. It is furthermore necessary to include mesonic fluctuations beyond the mean field for a quantitatively reliable description of the phase diagram.
\end{section}

\begin{acknowledgments}
This work is supported in part by BMBF and by the DFG Cluster of Excellence ``Origin and Structure of the Universe.'' One of the authors (W. W.) thanks Christof Wetterich for stimulating discussions.
\end{acknowledgments}

\bibliography{biblio}

\begin{thebibliography}{53}%
\makeatletter
\providecommand \@ifxundefined [1]{%
 \@ifx{#1\undefined}
}%
\providecommand \@ifnum [1]{%
 \ifnum #1\expandafter \@firstoftwo
 \else \expandafter \@secondoftwo
 \fi
}%
\providecommand \@ifx [1]{%
 \ifx #1\expandafter \@firstoftwo
 \else \expandafter \@secondoftwo
 \fi
}%
\providecommand \natexlab [1]{#1}%
\providecommand \enquote  [1]{``#1''}%
\providecommand \bibnamefont  [1]{#1}%
\providecommand \bibfnamefont [1]{#1}%
\providecommand \citenamefont [1]{#1}%
\providecommand \href@noop [0]{\@secondoftwo}%
\providecommand \href [0]{\begingroup \@sanitize@url \@href}%
\providecommand \@href[1]{\@@startlink{#1}\@@href}%
\providecommand \@@href[1]{\endgroup#1\@@endlink}%
\providecommand \@sanitize@url [0]{\catcode `\\12\catcode `\$12\catcode
  `\&12\catcode `\#12\catcode `\^12\catcode `\_12\catcode `\%12\relax}%
\providecommand \@@startlink[1]{}%
\providecommand \@@endlink[0]{}%
\providecommand \url  [0]{\begingroup\@sanitize@url \@url }%
\providecommand \@url [1]{\endgroup\@href {#1}{\urlprefix }}%
\providecommand \urlprefix  [0]{URL }%
\providecommand \Eprint [0]{\href }%
\providecommand \doibase [0]{http://dx.doi.org/}%
\providecommand \selectlanguage [0]{\@gobble}%
\providecommand \bibinfo  [0]{\@secondoftwo}%
\providecommand \bibfield  [0]{\@secondoftwo}%
\providecommand \translation [1]{[#1]}%
\providecommand \BibitemOpen [0]{}%
\providecommand \bibitemStop [0]{}%
\providecommand \bibitemNoStop [0]{.\EOS\space}%
\providecommand \EOS [0]{\spacefactor3000\relax}%
\providecommand \BibitemShut  [1]{\csname bibitem#1\endcsname}%
\let\auto@bib@innerbib\@empty
%</preamble>
\bibitem [{\citenamefont {Philipsen}(2005)}]{Philipsen2005qcd}%
  \BibitemOpen
  \bibfield  {author} {\bibinfo {author} {\bibfnamefont {O.}~\bibnamefont
  {Philipsen}},\ }\href@noop {} {\bibfield  {journal} {\bibinfo  {journal}
  {{PoS} {LAT2005}}\ }\textbf {\bibinfo {volume} {016}} (\bibinfo {year}
  {2005})},\ \Eprint {http://arxiv.org/abs/hep-lat/0510077}
  {{arXiv}:hep-lat/0510077} \BibitemShut {NoStop}%
\bibitem [{\citenamefont {Schmidt}(2006)}]{Schmidt2006lattice}%
  \BibitemOpen
  \bibfield  {author} {\bibinfo {author} {\bibfnamefont {C.}~\bibnamefont
  {Schmidt}},\ }\href@noop {} {\bibfield  {journal} {\bibinfo  {journal} {{PoS}
  {LAT2006}}\ }\textbf {\bibinfo {volume} {021}} (\bibinfo {year} {2006})},\
  \Eprint {http://arxiv.org/abs/hep-lat/0610116} {{arXiv}:hep-lat/0610116}
  \BibitemShut {NoStop}%
\bibitem [{\citenamefont {Philipsen}(2008)}]{Philipsen2008status}%
  \BibitemOpen
  \bibfield  {author} {\bibinfo {author} {\bibfnamefont {O.}~\bibnamefont
  {Philipsen}},\ }\href@noop {} {\bibfield  {journal} {\bibinfo  {journal}
  {Prog. Theor. Phys. Suppl.}\ }\textbf {\bibinfo {volume} {174}},\ \bibinfo
  {pages} {206} (\bibinfo {year} {2008})},\ \Eprint
  {http://arxiv.org/abs/0808.0672} {{arXiv}:0808.0672} \BibitemShut {NoStop}%
\bibitem [{\citenamefont {de~Forcrand}(2009)}]{de_Forcrand2009simulating}%
  \BibitemOpen
  \bibfield  {author} {\bibinfo {author} {\bibfnamefont {P.}~\bibnamefont
  {de~Forcrand}},\ }\href@noop {} {\bibfield  {journal} {\bibinfo  {journal}
  {{PoS} {LAT2009}}\ }\textbf {\bibinfo {volume} {010}} (\bibinfo {year}
  {2009})},\ \Eprint {http://arxiv.org/abs/1005.0539} {{arXiv}:1005.0539}
  \BibitemShut {NoStop}%
\bibitem [{\citenamefont {de~Forcrand}\ and\ \citenamefont
  {Philipsen}(2008)}]{forcrand2008chiral}%
  \BibitemOpen
  \bibfield  {author} {\bibinfo {author} {\bibfnamefont {P.}~\bibnamefont
  {de~Forcrand}}\ and\ \bibinfo {author} {\bibfnamefont {O.}~\bibnamefont
  {Philipsen}},\ }\href@noop {} {\bibfield  {journal} {\bibinfo  {journal}
  {{JHEP}}\ }\textbf {\bibinfo {volume} {0811}},\ \bibinfo {pages} {012}
  (\bibinfo {year} {2008})},\ \Eprint {http://arxiv.org/abs/0808.1096}
  {{arXiv}:0808.1096} \BibitemShut {NoStop}%
\bibitem [{\citenamefont {Fukushima}\ and\ \citenamefont
  {Sasaki}(2013)}]{fukushima2013phase}%
  \BibitemOpen
  \bibfield  {author} {\bibinfo {author} {\bibfnamefont {K.}~\bibnamefont
  {Fukushima}}\ and\ \bibinfo {author} {\bibfnamefont {C.}~\bibnamefont
  {Sasaki}},\ }\href@noop {} {\bibfield  {journal} {\bibinfo  {journal} {Prog.
  Part. Nucl. Phys.}\ }\textbf {\bibinfo {volume} {72}},\ \bibinfo {pages} {99}
  (\bibinfo {year} {2013})},\ \Eprint {http://arxiv.org/abs/1301.6377}
  {{arXiv}:1301.6377} \BibitemShut {NoStop}%
\bibitem [{\citenamefont {Fukushima}(2004)}]{fukushima2004chiral}%
  \BibitemOpen
  \bibfield  {author} {\bibinfo {author} {\bibfnamefont {K.}~\bibnamefont
  {Fukushima}},\ }\href@noop {} {\bibfield  {journal} {\bibinfo  {journal}
  {Phys. Lett. B}\ }\textbf {\bibinfo {volume} {591}},\ \bibinfo {pages} {277}
  (\bibinfo {year} {2004})},\ \Eprint {http://arxiv.org/abs/hep-ph/0310121}
  {{arXiv}:hep-ph/0310121} \BibitemShut {NoStop}%
\bibitem [{\citenamefont {R\"o{\ss}ner}\ \emph {et~al.}(2007)\citenamefont
  {R\"o{\ss}ner}, \citenamefont {Ratti},\ and\ \citenamefont
  {Weise}}]{roener2007polyakov}%
  \BibitemOpen
  \bibfield  {author} {\bibinfo {author} {\bibfnamefont {S.}~\bibnamefont
  {R\"o{\ss}ner}}, \bibinfo {author} {\bibfnamefont {C.}~\bibnamefont {Ratti}},
  \ and\ \bibinfo {author} {\bibfnamefont {W.}~\bibnamefont {Weise}},\
  }\href@noop {} {\bibfield  {journal} {\bibinfo  {journal} {Phys. Rev. D}\
  }\textbf {\bibinfo {volume} {75}},\ \bibinfo {pages} {034007} (\bibinfo
  {year} {2007})},\ \Eprint {http://arxiv.org/abs/hep-ph/0609281}
  {{arXiv}:hep-ph/0609281} \BibitemShut {NoStop}%
\bibitem [{\citenamefont {Ratti}\ \emph {et~al.}(2007)\citenamefont {Ratti},
  \citenamefont {R\"o{\ss}ner}, \citenamefont {Thaler},\ and\ \citenamefont
  {Weise}}]{ratti2006thermodynamics}%
  \BibitemOpen
  \bibfield  {author} {\bibinfo {author} {\bibfnamefont {C.}~\bibnamefont
  {Ratti}}, \bibinfo {author} {\bibfnamefont {S.}~\bibnamefont {R\"o{\ss}ner}},
  \bibinfo {author} {\bibfnamefont {M.}~\bibnamefont {Thaler}}, \ and\ \bibinfo
  {author} {\bibfnamefont {W.}~\bibnamefont {Weise}},\ }\href@noop {}
  {\bibfield  {journal} {\bibinfo  {journal} {Eur. Phys. J. C}\ }\textbf
  {\bibinfo {volume} {49}},\ \bibinfo {pages} {213} (\bibinfo {year} {2007})},\
  \Eprint {http://arxiv.org/abs/hep-ph/0609218} {{arXiv}:hep-ph/0609218}
  \BibitemShut {NoStop}%
\bibitem [{\citenamefont {Klimt}\ \emph {et~al.}(1990)\citenamefont {Klimt},
  \citenamefont {Lutz},\ and\ \citenamefont {Weise}}]{klimt1990chiral}%
  \BibitemOpen
  \bibfield  {author} {\bibinfo {author} {\bibfnamefont {S.}~\bibnamefont
  {Klimt}}, \bibinfo {author} {\bibfnamefont {M.}~\bibnamefont {Lutz}}, \ and\
  \bibinfo {author} {\bibfnamefont {W.}~\bibnamefont {Weise}},\ }\href@noop {}
  {\bibfield  {journal} {\bibinfo  {journal} {Phys. Lett. B}\ }\textbf
  {\bibinfo {volume} {249}},\ \bibinfo {pages} {386} (\bibinfo {year}
  {1990})}\BibitemShut {NoStop}%
\bibitem [{\citenamefont {Sasaki}\ \emph {et~al.}(2007)\citenamefont {Sasaki},
  \citenamefont {Friman},\ and\ \citenamefont {Redlich}}]{sasaki2007quark}%
  \BibitemOpen
  \bibfield  {author} {\bibinfo {author} {\bibfnamefont {C.}~\bibnamefont
  {Sasaki}}, \bibinfo {author} {\bibfnamefont {B.}~\bibnamefont {Friman}}, \
  and\ \bibinfo {author} {\bibfnamefont {K.}~\bibnamefont {Redlich}},\
  }\href@noop {} {\bibfield  {journal} {\bibinfo  {journal} {Phys. Rev. D}\
  }\textbf {\bibinfo {volume} {75}},\ \bibinfo {pages} {054026} (\bibinfo
  {year} {2007})},\ \Eprint {http://arxiv.org/abs/hep-ph/0611143}
  {{arXiv}:hep-ph/0611143} \BibitemShut {NoStop}%
\bibitem [{\citenamefont {Kitazawa}\ \emph {et~al.}(2002)\citenamefont
  {Kitazawa}, \citenamefont {Koide}, \citenamefont {Kunihiro},\ and\
  \citenamefont {Nemoto}}]{kitazawa2002chiral}%
  \BibitemOpen
  \bibfield  {author} {\bibinfo {author} {\bibfnamefont {M.}~\bibnamefont
  {Kitazawa}}, \bibinfo {author} {\bibfnamefont {T.}~\bibnamefont {Koide}},
  \bibinfo {author} {\bibfnamefont {T.}~\bibnamefont {Kunihiro}}, \ and\
  \bibinfo {author} {\bibfnamefont {Y.}~\bibnamefont {Nemoto}},\ }\href@noop {}
  {\bibfield  {journal} {\bibinfo  {journal} {Prog. Theor. Phys.}\ }\textbf
  {\bibinfo {volume} {108}},\ \bibinfo {pages} {929} (\bibinfo {year}
  {2002})},\ \Eprint {http://arxiv.org/abs/hep-ph/0207255}
  {{arXiv}:hep-ph/0207255} \BibitemShut {NoStop}%
\bibitem [{\citenamefont {Fukushima}(2008{\natexlab{a}})}]{fukushima2008phase}%
  \BibitemOpen
  \bibfield  {author} {\bibinfo {author} {\bibfnamefont {K.}~\bibnamefont
  {Fukushima}},\ }\href@noop {} {\bibfield  {journal} {\bibinfo  {journal}
  {Phys. Rev. D}\ }\textbf {\bibinfo {volume} {77}},\ \bibinfo {pages} {114028}
  (\bibinfo {year} {2008}{\natexlab{a}})},\ \Eprint
  {http://arxiv.org/abs/0803.3318} {{arXiv}:0803.3318} \BibitemShut {NoStop}%
\bibitem [{\citenamefont {Bratovic}\ \emph {et~al.}(2013)\citenamefont
  {Bratovic}, \citenamefont {Hatsuda},\ and\ \citenamefont
  {Weise}}]{bratovic2013role}%
  \BibitemOpen
  \bibfield  {author} {\bibinfo {author} {\bibfnamefont {N.~M.}\ \bibnamefont
  {Bratovic}}, \bibinfo {author} {\bibfnamefont {T.}~\bibnamefont {Hatsuda}}, \
  and\ \bibinfo {author} {\bibfnamefont {W.}~\bibnamefont {Weise}},\
  }\href@noop {} {\bibfield  {journal} {\bibinfo  {journal} {Phys. Lett. B}\
  }\textbf {\bibinfo {volume} {719}},\ \bibinfo {pages} {131} (\bibinfo {year}
  {2013})},\ \Eprint {http://arxiv.org/abs/1204.3788} {{arXiv}:1204.3788}
  \BibitemShut {NoStop}%
\bibitem [{\citenamefont {Hell}\ \emph {et~al.}(2013)\citenamefont {Hell},
  \citenamefont {Kashiwa},\ and\ \citenamefont {Weise}}]{hell2013impact}%
  \BibitemOpen
  \bibfield  {author} {\bibinfo {author} {\bibfnamefont {T.}~\bibnamefont
  {Hell}}, \bibinfo {author} {\bibfnamefont {K.}~\bibnamefont {Kashiwa}}, \
  and\ \bibinfo {author} {\bibfnamefont {W.}~\bibnamefont {Weise}},\
  }\href@noop {} {\bibfield  {journal} {\bibinfo  {journal} {Jour. Mod. Phys.}\
  }\textbf {\bibinfo {volume} {04}},\ \bibinfo {pages} {644} (\bibinfo {year}
  {2013})},\ \Eprint {http://arxiv.org/abs/1212.4017} {{arXiv}:1212.4017}
  \BibitemShut {NoStop}%
\bibitem [{\citenamefont
  {Fukushima}(2008{\natexlab{b}})}]{fukushima2008critical}%
  \BibitemOpen
  \bibfield  {author} {\bibinfo {author} {\bibfnamefont {K.}~\bibnamefont
  {Fukushima}},\ }\href@noop {} {\bibfield  {journal} {\bibinfo  {journal}
  {Phys. Rev. D}\ }\textbf {\bibinfo {volume} {78}},\ \bibinfo {pages} {114019}
  (\bibinfo {year} {2008}{\natexlab{b}})},\ \Eprint
  {http://arxiv.org/abs/0809.3080} {{arXiv}:0809.3080} \BibitemShut {NoStop}%
\bibitem [{\citenamefont {Schaefer}\ and\ \citenamefont
  {Wagner}(2012)}]{schaefer2012qcd}%
  \BibitemOpen
  \bibfield  {author} {\bibinfo {author} {\bibfnamefont {B.~J.}\ \bibnamefont
  {Schaefer}}\ and\ \bibinfo {author} {\bibfnamefont {M.}~\bibnamefont
  {Wagner}},\ }\href@noop {} {\bibfield  {journal} {\bibinfo  {journal} {Phys.
  Rev. D}\ }\textbf {\bibinfo {volume} {85}},\ \bibinfo {pages} {034027}
  (\bibinfo {year} {2012})},\ \Eprint {http://arxiv.org/abs/1111.6871}
  {{arXiv}:1111.6871} \BibitemShut {NoStop}%
\bibitem [{\citenamefont {Schaefer}\ \emph {et~al.}(2007)\citenamefont
  {Schaefer}, \citenamefont {Pawlowski},\ and\ \citenamefont
  {Wambach}}]{schaefer2007phase}%
  \BibitemOpen
  \bibfield  {author} {\bibinfo {author} {\bibfnamefont {B.~J.}\ \bibnamefont
  {Schaefer}}, \bibinfo {author} {\bibfnamefont {J.~M.}\ \bibnamefont
  {Pawlowski}}, \ and\ \bibinfo {author} {\bibfnamefont {J.}~\bibnamefont
  {Wambach}},\ }\href@noop {} {\bibfield  {journal} {\bibinfo  {journal} {Phys.
  Rev. D}\ }\textbf {\bibinfo {volume} {76}},\ \bibinfo {pages} {074023}
  (\bibinfo {year} {2007})},\ \Eprint {http://arxiv.org/abs/0704.3234}
  {{arXiv}:0704.3234} \BibitemShut {NoStop}%
\bibitem [{\citenamefont {Skokov}\ \emph
  {et~al.}(2010{\natexlab{a}})\citenamefont {Skokov}, \citenamefont {Stokic},
  \citenamefont {Friman},\ and\ \citenamefont {Redlich}}]{Skokov2010Meson}%
  \BibitemOpen
  \bibfield  {author} {\bibinfo {author} {\bibfnamefont {V.}~\bibnamefont
  {Skokov}}, \bibinfo {author} {\bibfnamefont {B.}~\bibnamefont {Stokic}},
  \bibinfo {author} {\bibfnamefont {B.}~\bibnamefont {Friman}}, \ and\ \bibinfo
  {author} {\bibfnamefont {K.}~\bibnamefont {Redlich}},\ }\href@noop {}
  {\bibfield  {journal} {\bibinfo  {journal} {Phys. Rev. C}\ }\textbf {\bibinfo
  {volume} {82}},\ \bibinfo {pages} {015206} (\bibinfo {year}
  {2010}{\natexlab{a}})},\ \Eprint {http://arxiv.org/abs/1004.2665}
  {{arXiv}:1004.2665} \BibitemShut {NoStop}%
\bibitem [{\citenamefont {Skokov}\ \emph {et~al.}(2011)\citenamefont {Skokov},
  \citenamefont {Friman},\ and\ \citenamefont {Redlich}}]{Skokov2011Quark}%
  \BibitemOpen
  \bibfield  {author} {\bibinfo {author} {\bibfnamefont {V.}~\bibnamefont
  {Skokov}}, \bibinfo {author} {\bibfnamefont {B.}~\bibnamefont {Friman}}, \
  and\ \bibinfo {author} {\bibfnamefont {K.}~\bibnamefont {Redlich}},\
  }\href@noop {} {\bibfield  {journal} {\bibinfo  {journal} {Phys. Rev. C}\
  }\textbf {\bibinfo {volume} {83}},\ \bibinfo {pages} {054904} (\bibinfo
  {year} {2011})},\ \Eprint {http://arxiv.org/abs/1008.4570}
  {{arXiv}:1008.4570} \BibitemShut {NoStop}%
\bibitem [{\citenamefont {Herbst}\ \emph {et~al.}(2011)\citenamefont {Herbst},
  \citenamefont {Pawlowski},\ and\ \citenamefont {Schaefer}}]{Herbst2011Phase}%
  \BibitemOpen
  \bibfield  {author} {\bibinfo {author} {\bibfnamefont {T.~K.}\ \bibnamefont
  {Herbst}}, \bibinfo {author} {\bibfnamefont {J.~M.}\ \bibnamefont
  {Pawlowski}}, \ and\ \bibinfo {author} {\bibfnamefont {B.~J.}\ \bibnamefont
  {Schaefer}},\ }\href@noop {} {\bibfield  {journal} {\bibinfo  {journal}
  {Phys. Lett. B}\ }\textbf {\bibinfo {volume} {696}},\ \bibinfo {pages} {58}
  (\bibinfo {year} {2011})},\ \Eprint {http://arxiv.org/abs/1008.0081}
  {{arXiv}:1008.0081} \BibitemShut {NoStop}%
\bibitem [{\citenamefont {Herbst}\ \emph {et~al.}(2013)\citenamefont {Herbst},
  \citenamefont {Pawlowski},\ and\ \citenamefont {Schaefer}}]{Herbst2013Phase}%
  \BibitemOpen
  \bibfield  {author} {\bibinfo {author} {\bibfnamefont {T.~K.}\ \bibnamefont
  {Herbst}}, \bibinfo {author} {\bibfnamefont {J.~M.}\ \bibnamefont
  {Pawlowski}}, \ and\ \bibinfo {author} {\bibfnamefont {B.~J.}\ \bibnamefont
  {Schaefer}},\ }\href@noop {} {\bibfield  {journal} {\bibinfo  {journal}
  {Phys. Rev. D}\ }\textbf {\bibinfo {volume} {88}},\ \bibinfo {pages} {014007}
  (\bibinfo {year} {2013})},\ \Eprint {http://arxiv.org/abs/1302.1426}
  {{arXiv}:1302.1426} \BibitemShut {NoStop}%
\bibitem [{\citenamefont {Fischer}\ and\ \citenamefont
  {Luecker}(2013)}]{fischer2012propagators}%
  \BibitemOpen
  \bibfield  {author} {\bibinfo {author} {\bibfnamefont {C.~S.}\ \bibnamefont
  {Fischer}}\ and\ \bibinfo {author} {\bibfnamefont {J.}~\bibnamefont
  {Luecker}},\ }\href@noop {} {\bibfield  {journal} {\bibinfo  {journal} {Phys.
  Lett. B}\ }\textbf {\bibinfo {volume} {718}},\ \bibinfo {pages} {1036}
  (\bibinfo {year} {2013})},\ \Eprint {http://arxiv.org/abs/1206.5191}
  {{arXiv}:1206.5191} \BibitemShut {NoStop}%
\bibitem [{\citenamefont {Fischer}\ \emph {et~al.}(2013)\citenamefont
  {Fischer}, \citenamefont {Fister}, \citenamefont {Luecker},\ and\
  \citenamefont {Pawlowski}}]{fischer2013polyakov}%
  \BibitemOpen
  \bibfield  {author} {\bibinfo {author} {\bibfnamefont {C.~S.}\ \bibnamefont
  {Fischer}}, \bibinfo {author} {\bibfnamefont {L.}~\bibnamefont {Fister}},
  \bibinfo {author} {\bibfnamefont {J.}~\bibnamefont {Luecker}}, \ and\
  \bibinfo {author} {\bibfnamefont {J.~M.}\ \bibnamefont {Pawlowski}},\
  }\href@noop {} {\  (\bibinfo {year} {2013})},\ \Eprint
  {http://arxiv.org/abs/1306.6022} {{arXiv}:1306.6022} \BibitemShut {NoStop}%
\bibitem [{\citenamefont {Nickel}(2009)}]{nickel_how_2009}%
  \BibitemOpen
  \bibfield  {author} {\bibinfo {author} {\bibfnamefont {D.}~\bibnamefont
  {Nickel}},\ }\href@noop {} {\bibfield  {journal} {\bibinfo  {journal} {Phys.
  Rev. Lett.}\ }\textbf {\bibinfo {volume} {103}},\ \bibinfo {pages} {072301}
  (\bibinfo {year} {2009})},\ \Eprint {http://arxiv.org/abs/0902.1778}
  {{arXiv}:0902.1778} \BibitemShut {NoStop}%
\bibitem [{\citenamefont {Ooguri}\ and\ \citenamefont
  {Park}(2011)}]{ooguri_spatially_2011}%
  \BibitemOpen
  \bibfield  {author} {\bibinfo {author} {\bibfnamefont {H.}~\bibnamefont
  {Ooguri}}\ and\ \bibinfo {author} {\bibfnamefont {C.-S.}\ \bibnamefont
  {Park}},\ }\href@noop {} {\bibfield  {journal} {\bibinfo  {journal} {Phys.
  Rev. Lett.}\ }\textbf {\bibinfo {volume} {106}},\ \bibinfo {pages} {061601}
  (\bibinfo {year} {2011})},\ \Eprint {http://arxiv.org/abs/1011.4144}
  {{arXiv}:1011.4144} \BibitemShut {NoStop}%
\bibitem [{\citenamefont {Fukushima}\ and\ \citenamefont
  {Morales}(2013)}]{fukushima_spatial_2013}%
  \BibitemOpen
  \bibfield  {author} {\bibinfo {author} {\bibfnamefont {K.}~\bibnamefont
  {Fukushima}}\ and\ \bibinfo {author} {\bibfnamefont {P.~A.}\ \bibnamefont
  {Morales}},\ }\href@noop {} {\bibfield  {journal} {\bibinfo  {journal} {Phys.
  Rev. Lett.}\ }\textbf {\bibinfo {volume} {111}},\ \bibinfo {pages} {051601}
  (\bibinfo {year} {2013})},\ \Eprint {http://arxiv.org/abs/1305.4115}
  {{arXiv}:1305.4115} \BibitemShut {NoStop}%
\bibitem [{\citenamefont {Berges}\ \emph {et~al.}(2003)\citenamefont {Berges},
  \citenamefont {Jungnickel},\ and\ \citenamefont
  {Wetterich}}]{berges2003quark}%
  \BibitemOpen
  \bibfield  {author} {\bibinfo {author} {\bibfnamefont {J.}~\bibnamefont
  {Berges}}, \bibinfo {author} {\bibfnamefont {D.}~\bibnamefont {Jungnickel}},
  \ and\ \bibinfo {author} {\bibfnamefont {C.}~\bibnamefont {Wetterich}},\
  }\href@noop {} {\bibfield  {journal} {\bibinfo  {journal} {Int. J. Mod. Phys.
  A}\ }\textbf {\bibinfo {volume} {18}},\ \bibinfo {pages} {3189} (\bibinfo
  {year} {2003})},\ \Eprint {http://arxiv.org/abs/hep-ph/9811387}
  {{arXiv}:hep-ph/9811387} \BibitemShut {NoStop}%
\bibitem [{\citenamefont {Floerchinger}\ and\ \citenamefont
  {Wetterich}(2012)}]{floerchinger2012chemical}%
  \BibitemOpen
  \bibfield  {author} {\bibinfo {author} {\bibfnamefont {S.}~\bibnamefont
  {Floerchinger}}\ and\ \bibinfo {author} {\bibfnamefont {C.}~\bibnamefont
  {Wetterich}},\ }\href@noop {} {\bibfield  {journal} {\bibinfo  {journal}
  {Nucl. Phys. A}\ }\textbf {\bibinfo {volume} {890-891}},\ \bibinfo {pages}
  {11} (\bibinfo {year} {2012})},\ \Eprint {http://arxiv.org/abs/1202.1671}
  {{arXiv}:1202.1671} \BibitemShut {NoStop}%
\bibitem [{\citenamefont {{Braun-Munzinger}}\ \emph {et~al.}(2004)\citenamefont
  {{Braun-Munzinger}}, \citenamefont {Stachel},\ and\ \citenamefont
  {Wetterich}}]{Braun-Munzinger2004Chemical}%
  \BibitemOpen
  \bibfield  {author} {\bibinfo {author} {\bibfnamefont {P.}~\bibnamefont
  {{Braun-Munzinger}}}, \bibinfo {author} {\bibfnamefont {J.}~\bibnamefont
  {Stachel}}, \ and\ \bibinfo {author} {\bibfnamefont {C.}~\bibnamefont
  {Wetterich}},\ }\href@noop {} {\bibfield  {journal} {\bibinfo  {journal}
  {Phys. Lett. B}\ }\textbf {\bibinfo {volume} {596}},\ \bibinfo {pages} {61}
  (\bibinfo {year} {2004})},\ \Eprint {http://arxiv.org/abs/nucl-th/0311005}
  {{arXiv}:nucl-th/0311005} \BibitemShut {NoStop}%
\bibitem [{\citenamefont {{Kaczmarek \textit{et
  al.}}}(2011)}]{Kaczmarek2011Phase}%
  \BibitemOpen
  \bibfield  {author} {\bibinfo {author} {\bibfnamefont {O.}~\bibnamefont
  {{Kaczmarek \textit{et al.}}}},\ }\href@noop {} {\bibfield  {journal}
  {\bibinfo  {journal} {Phys. Rev. D}\ }\textbf {\bibinfo {volume} {83}},\
  \bibinfo {pages} {014504} (\bibinfo {year} {2011})},\ \Eprint
  {http://arxiv.org/abs/1011.3130} {{arXiv}:1011.3130} \BibitemShut {NoStop}%
\bibitem [{\citenamefont {Karsch}(2012)}]{Karsch2012Determination}%
  \BibitemOpen
  \bibfield  {author} {\bibinfo {author} {\bibfnamefont {F.}~\bibnamefont
  {Karsch}},\ }\href@noop {} {\bibfield  {journal} {\bibinfo  {journal} {Centr.
  Eur. J. Phys.}\ }\textbf {\bibinfo {volume} {10}},\ \bibinfo {pages} {1234}
  (\bibinfo {year} {2012})},\ \Eprint {http://arxiv.org/abs/1202.4173}
  {{arXiv}:1202.4173} \BibitemShut {NoStop}%
\bibitem [{\citenamefont {Wetterich}(1993)}]{wetterich1993exact}%
  \BibitemOpen
  \bibfield  {author} {\bibinfo {author} {\bibfnamefont {C.}~\bibnamefont
  {Wetterich}},\ }\href@noop {} {\bibfield  {journal} {\bibinfo  {journal}
  {Phys. Lett. B}\ }\textbf {\bibinfo {volume} {301}},\ \bibinfo {pages} {90}
  (\bibinfo {year} {1993})}\BibitemShut {NoStop}%
\bibitem [{\citenamefont {Fiorilla}\ \emph
  {et~al.}(2012{\natexlab{a}})\citenamefont {Fiorilla}, \citenamefont
  {Kaiser},\ and\ \citenamefont {Weise}}]{fiorilla2012chiral}%
  \BibitemOpen
  \bibfield  {author} {\bibinfo {author} {\bibfnamefont {S.}~\bibnamefont
  {Fiorilla}}, \bibinfo {author} {\bibfnamefont {N.}~\bibnamefont {Kaiser}}, \
  and\ \bibinfo {author} {\bibfnamefont {W.}~\bibnamefont {Weise}},\
  }\href@noop {} {\bibfield  {journal} {\bibinfo  {journal} {Nucl. Phys. A}\
  }\textbf {\bibinfo {volume} {880}},\ \bibinfo {pages} {65} (\bibinfo {year}
  {2012}{\natexlab{a}})},\ \Eprint {http://arxiv.org/abs/1111.2791}
  {{arXiv}:1111.2791} \BibitemShut {NoStop}%
\bibitem [{\citenamefont {Fiorilla}\ \emph
  {et~al.}(2012{\natexlab{b}})\citenamefont {Fiorilla}, \citenamefont
  {Kaiser},\ and\ \citenamefont {Weise}}]{fiorilla2012nuclear}%
  \BibitemOpen
  \bibfield  {author} {\bibinfo {author} {\bibfnamefont {S.}~\bibnamefont
  {Fiorilla}}, \bibinfo {author} {\bibfnamefont {N.}~\bibnamefont {Kaiser}}, \
  and\ \bibinfo {author} {\bibfnamefont {W.}~\bibnamefont {Weise}},\
  }\href@noop {} {\bibfield  {journal} {\bibinfo  {journal} {Phys. Lett. B}\
  }\textbf {\bibinfo {volume} {714}},\ \bibinfo {pages} {251} (\bibinfo {year}
  {2012}{\natexlab{b}})},\ \Eprint {http://arxiv.org/abs/1204.4318}
  {{arXiv}:1204.4318} \BibitemShut {NoStop}%
\bibitem [{\citenamefont {Walecka}(1974)}]{walecka1974theory}%
  \BibitemOpen
  \bibfield  {author} {\bibinfo {author} {\bibfnamefont {J.}~\bibnamefont
  {Walecka}},\ }\href@noop {} {\bibfield  {journal} {\bibinfo  {journal} {Ann.
  Phys.}\ }\textbf {\bibinfo {volume} {83}},\ \bibinfo {pages} {491} (\bibinfo
  {year} {1974})}\BibitemShut {NoStop}%
\bibitem [{\citenamefont {Berges}\ \emph {et~al.}(2000)\citenamefont {Berges},
  \citenamefont {Jungnickel},\ and\ \citenamefont
  {Wetterich}}]{berges2000chiral}%
  \BibitemOpen
  \bibfield  {author} {\bibinfo {author} {\bibfnamefont {J.}~\bibnamefont
  {Berges}}, \bibinfo {author} {\bibfnamefont {D.~U.}\ \bibnamefont
  {Jungnickel}}, \ and\ \bibinfo {author} {\bibfnamefont {C.}~\bibnamefont
  {Wetterich}},\ }\href@noop {} {\bibfield  {journal} {\bibinfo  {journal}
  {Eur. Phys. J. C}\ }\textbf {\bibinfo {volume} {13}},\ \bibinfo {pages} {323}
  (\bibinfo {year} {2000})},\ \Eprint {http://arxiv.org/abs/hep-ph/9811347}
  {{arXiv}:hep-ph/9811347} \BibitemShut {NoStop}%
\bibitem [{\citenamefont {Litim}(2000)}]{Litim2000optimisation}%
  \BibitemOpen
  \bibfield  {author} {\bibinfo {author} {\bibfnamefont {D.~F.}\ \bibnamefont
  {Litim}},\ }\href@noop {} {\bibfield  {journal} {\bibinfo  {journal} {Phys.
  Lett. B}\ }\textbf {\bibinfo {volume} {486}},\ \bibinfo {pages} {92}
  (\bibinfo {year} {2000})},\ \Eprint {http://arxiv.org/abs/hep-th/0005245}
  {{arXiv}:hep-th/0005245} \BibitemShut {NoStop}%
\bibitem [{\citenamefont {Litim}(2001{\natexlab{a}})}]{litim2001optimised}%
  \BibitemOpen
  \bibfield  {author} {\bibinfo {author} {\bibfnamefont {D.~F.}\ \bibnamefont
  {Litim}},\ }\href@noop {} {\bibfield  {journal} {\bibinfo  {journal} {Phys.
  Rev. D}\ }\textbf {\bibinfo {volume} {64}},\ \bibinfo {pages} {105007}
  (\bibinfo {year} {2001}{\natexlab{a}})},\ \Eprint
  {http://arxiv.org/abs/hep-th/0103195} {{arXiv}:hep-th/0103195} \BibitemShut
  {NoStop}%
\bibitem [{\citenamefont {Litim}(2001{\natexlab{b}})}]{litim2001mind}%
  \BibitemOpen
  \bibfield  {author} {\bibinfo {author} {\bibfnamefont {D.~F.}\ \bibnamefont
  {Litim}},\ }\href@noop {} {\bibfield  {journal} {\bibinfo  {journal} {Int. J.
  Mod. Phys. A}\ }\textbf {\bibinfo {volume} {16}},\ \bibinfo {pages} {2081}
  (\bibinfo {year} {2001}{\natexlab{b}})},\ \Eprint
  {http://arxiv.org/abs/hep-th/0104221} {{arXiv}:hep-th/0104221} \BibitemShut
  {NoStop}%
\bibitem [{\citenamefont {Pawlowski}(2007)}]{pawlowski2007aspects}%
  \BibitemOpen
  \bibfield  {author} {\bibinfo {author} {\bibfnamefont {J.~M.}\ \bibnamefont
  {Pawlowski}},\ }\href@noop {} {\bibfield  {journal} {\bibinfo  {journal}
  {Ann. Phys.}\ }\textbf {\bibinfo {volume} {322}},\ \bibinfo {pages} {2831}
  (\bibinfo {year} {2007})},\ \Eprint {http://arxiv.org/abs/hep-th/0512261}
  {{arXiv}:hep-th/0512261} \BibitemShut {NoStop}%
\bibitem [{\citenamefont {Litim}\ and\ \citenamefont
  {Pawlowski}(2006)}]{litim2006non-perturbative}%
  \BibitemOpen
  \bibfield  {author} {\bibinfo {author} {\bibfnamefont {D.~F.}\ \bibnamefont
  {Litim}}\ and\ \bibinfo {author} {\bibfnamefont {J.~M.}\ \bibnamefont
  {Pawlowski}},\ }\href@noop {} {\bibfield  {journal} {\bibinfo  {journal}
  {{JHEP}}\ }\textbf {\bibinfo {volume} {0611}},\ \bibinfo {pages} {026}
  (\bibinfo {year} {2006})},\ \Eprint {http://arxiv.org/abs/hep-th/0609122}
  {{arXiv}:hep-th/0609122} \BibitemShut {NoStop}%
\bibitem [{\citenamefont {Blaizot}\ \emph {et~al.}(2007)\citenamefont
  {Blaizot}, \citenamefont {Ipp}, \citenamefont {{Mendez-Galain}},\ and\
  \citenamefont {Wschebor}}]{blaizot2007perturbation}%
  \BibitemOpen
  \bibfield  {author} {\bibinfo {author} {\bibfnamefont {J.-P.}\ \bibnamefont
  {Blaizot}}, \bibinfo {author} {\bibfnamefont {A.}~\bibnamefont {Ipp}},
  \bibinfo {author} {\bibfnamefont {R.}~\bibnamefont {{Mendez-Galain}}}, \ and\
  \bibinfo {author} {\bibfnamefont {N.}~\bibnamefont {Wschebor}},\ }\href@noop
  {} {\bibfield  {journal} {\bibinfo  {journal} {Nucl. Phys. A}\ }\textbf
  {\bibinfo {volume} {784}},\ \bibinfo {pages} {376} (\bibinfo {year}
  {2007})},\ \Eprint {http://arxiv.org/abs/hep-ph/0610004}
  {{arXiv}:hep-ph/0610004} \BibitemShut {NoStop}%
\bibitem [{\citenamefont {Braun}\ \emph {et~al.}(2004)\citenamefont {Braun},
  \citenamefont {Pirner},\ and\ \citenamefont {Schwenzer}}]{braun2003linking}%
  \BibitemOpen
  \bibfield  {author} {\bibinfo {author} {\bibfnamefont {J.}~\bibnamefont
  {Braun}}, \bibinfo {author} {\bibfnamefont {H.}~\bibnamefont {Pirner}}, \
  and\ \bibinfo {author} {\bibfnamefont {K.}~\bibnamefont {Schwenzer}},\
  }\href@noop {} {\bibfield  {journal} {\bibinfo  {journal} {Phys. Rev. D}\
  }\textbf {\bibinfo {volume} {70}},\ \bibinfo {pages} {085016} (\bibinfo
  {year} {2004})},\ \Eprint {http://arxiv.org/abs/hep-ph/0312277}
  {{arXiv}:hep-ph/0312277} \BibitemShut {NoStop}%
\bibitem [{\citenamefont {Schaefer}\ and\ \citenamefont
  {Wambach}(2005)}]{schaefer2005phase}%
  \BibitemOpen
  \bibfield  {author} {\bibinfo {author} {\bibfnamefont {B.~J.}\ \bibnamefont
  {Schaefer}}\ and\ \bibinfo {author} {\bibfnamefont {J.}~\bibnamefont
  {Wambach}},\ }\href@noop {} {\bibfield  {journal} {\bibinfo  {journal} {Nucl.
  Phys. A}\ }\textbf {\bibinfo {volume} {757}},\ \bibinfo {pages} {479}
  (\bibinfo {year} {2005})},\ \Eprint {http://arxiv.org/abs/nucl-th/0403039}
  {{arXiv}:nucl-th/0403039} \BibitemShut {NoStop}%
\bibitem [{\citenamefont {Caprini}\ \emph {et~al.}(2006)\citenamefont
  {Caprini}, \citenamefont {Colangelo},\ and\ \citenamefont
  {Leutwyler}}]{caprini2006mass}%
  \BibitemOpen
  \bibfield  {author} {\bibinfo {author} {\bibfnamefont {I.}~\bibnamefont
  {Caprini}}, \bibinfo {author} {\bibfnamefont {G.}~\bibnamefont {Colangelo}},
  \ and\ \bibinfo {author} {\bibfnamefont {H.}~\bibnamefont {Leutwyler}},\
  }\href@noop {} {\bibfield  {journal} {\bibinfo  {journal} {Phys. Rev. Lett.}\
  }\textbf {\bibinfo {volume} {96}},\ \bibinfo {pages} {132001} (\bibinfo
  {year} {2006})},\ \Eprint {http://arxiv.org/abs/hep-ph/0512364}
  {{arXiv}:hep-ph/0512364} \BibitemShut {NoStop}%
\bibitem [{\citenamefont {{Garcia-Martin}}\ \emph {et~al.}(2007)\citenamefont
  {{Garcia-Martin}}, \citenamefont {Pelaez},\ and\ \citenamefont
  {Yndurain}}]{yndurain2007experimental}%
  \BibitemOpen
  \bibfield  {author} {\bibinfo {author} {\bibfnamefont {R.}~\bibnamefont
  {{Garcia-Martin}}}, \bibinfo {author} {\bibfnamefont {J.~R.}\ \bibnamefont
  {Pelaez}}, \ and\ \bibinfo {author} {\bibfnamefont {F.~J.}\ \bibnamefont
  {Yndurain}},\ }\href@noop {} {\bibfield  {journal} {\bibinfo  {journal}
  {Phys. Rev. D}\ }\textbf {\bibinfo {volume} {76}},\ \bibinfo {pages} {074034}
  (\bibinfo {year} {2007})},\ \Eprint {http://arxiv.org/abs/hep-ph/0701025}
  {{arXiv}:hep-ph/0701025} \BibitemShut {NoStop}%
\bibitem [{\citenamefont {{Adams \textit{et al.}}}(1995)}]{Adams1995Solving}%
  \BibitemOpen
  \bibfield  {author} {\bibinfo {author} {\bibfnamefont {J.}~\bibnamefont
  {{Adams \textit{et al.}}}},\ }\href@noop {} {\bibfield  {journal} {\bibinfo
  {journal} {Mod. Phys. Lett. A}\ }\textbf {\bibinfo {volume} {10}},\ \bibinfo
  {pages} {2367} (\bibinfo {year} {1995})},\ \Eprint
  {http://arxiv.org/abs/hep-th/9507093} {{arXiv}:hep-th/9507093} \BibitemShut
  {NoStop}%
\bibitem [{\citenamefont {Holt}\ \emph {et~al.}(2013)\citenamefont {Holt},
  \citenamefont {Kaiser},\ and\ \citenamefont {Weise}}]{Holt2013Nuclear}%
  \BibitemOpen
  \bibfield  {author} {\bibinfo {author} {\bibfnamefont {J.~W.}\ \bibnamefont
  {Holt}}, \bibinfo {author} {\bibfnamefont {N.}~\bibnamefont {Kaiser}}, \ and\
  \bibinfo {author} {\bibfnamefont {W.}~\bibnamefont {Weise}},\ }\href@noop {}
  {\bibfield  {journal} {\bibinfo  {journal} {Prog. Part. Nucl. Phys.}\
  }\textbf {\bibinfo {volume} {73}},\ \bibinfo {pages} {35} (\bibinfo {year}
  {2013})},\ \Eprint {http://arxiv.org/abs/1304.6350} {{arXiv}:1304.6350}
  \BibitemShut {NoStop}%
\bibitem [{\citenamefont {{Karnaukhov \textit{et
  al.}}}(2008)}]{Karnaukhov2008Critical}%
  \BibitemOpen
  \bibfield  {author} {\bibinfo {author} {\bibfnamefont {V.~A.}\ \bibnamefont
  {{Karnaukhov \textit{et al.}}}},\ }\href@noop {} {\bibfield  {journal}
  {\bibinfo  {journal} {Phys. Atom. Nuclei}\ }\textbf {\bibinfo {volume}
  {71}},\ \bibinfo {pages} {2067} (\bibinfo {year} {2008})},\ \Eprint
  {http://arxiv.org/abs/0801.4485} {{arXiv}:0801.4485} \BibitemShut {NoStop}%
\bibitem [{\citenamefont {Andronic}\ \emph {et~al.}(2009)\citenamefont
  {Andronic}, \citenamefont {{Braun-Munzinger}},\ and\ \citenamefont
  {Stachel}}]{Andronic2009Thermal}%
  \BibitemOpen
  \bibfield  {author} {\bibinfo {author} {\bibfnamefont {A.}~\bibnamefont
  {Andronic}}, \bibinfo {author} {\bibfnamefont {P.}~\bibnamefont
  {{Braun-Munzinger}}}, \ and\ \bibinfo {author} {\bibfnamefont
  {J.}~\bibnamefont {Stachel}},\ }\href@noop {} {\bibfield  {journal} {\bibinfo
   {journal} {Phys. Lett. B}\ }\textbf {\bibinfo {volume} {673}},\ \bibinfo
  {pages} {142} (\bibinfo {year} {2009})},\ \Eprint
  {http://arxiv.org/abs/0812.1186} {{arXiv}:0812.1186} \BibitemShut {NoStop}%
\bibitem [{\citenamefont {Schaefer}\ and\ \citenamefont
  {Wambach}(2007)}]{schaefer2007susceptibilities}%
  \BibitemOpen
  \bibfield  {author} {\bibinfo {author} {\bibfnamefont {B.~J.}\ \bibnamefont
  {Schaefer}}\ and\ \bibinfo {author} {\bibfnamefont {J.}~\bibnamefont
  {Wambach}},\ }\href@noop {} {\bibfield  {journal} {\bibinfo  {journal} {Phys.
  Rev. D}\ }\textbf {\bibinfo {volume} {75}},\ \bibinfo {pages} {085015}
  (\bibinfo {year} {2007})},\ \Eprint {http://arxiv.org/abs/hep-ph/0603256}
  {{arXiv}:hep-ph/0603256} \BibitemShut {NoStop}%
\bibitem [{\citenamefont {Skokov}\ \emph
  {et~al.}(2010{\natexlab{b}})\citenamefont {Skokov}, \citenamefont {Friman},
  \citenamefont {Nakano}, \citenamefont {Redlich},\ and\ \citenamefont
  {Schaefer}}]{skokov2010vacuum}%
  \BibitemOpen
  \bibfield  {author} {\bibinfo {author} {\bibfnamefont {V.}~\bibnamefont
  {Skokov}}, \bibinfo {author} {\bibfnamefont {B.}~\bibnamefont {Friman}},
  \bibinfo {author} {\bibfnamefont {E.}~\bibnamefont {Nakano}}, \bibinfo
  {author} {\bibfnamefont {K.}~\bibnamefont {Redlich}}, \ and\ \bibinfo
  {author} {\bibfnamefont {B.~J.}\ \bibnamefont {Schaefer}},\ }\href@noop {}
  {\bibfield  {journal} {\bibinfo  {journal} {Phys. Rev. D}\ }\textbf {\bibinfo
  {volume} {82}},\ \bibinfo {pages} {034029} (\bibinfo {year}
  {2010}{\natexlab{b}})},\ \Eprint {http://arxiv.org/abs/1005.3166}
  {{arXiv}:1005.3166} \BibitemShut {NoStop}%
\end{thebibliography}%
%\begin{bibliography}
%\input{NucleonMeson23082013.bbl}
%\end{bibliography}

\end{document}